\documentclass[twocolumn,preprintnumbers,amsmath,amssymb,superscriptaddress]{revtex4}
\usepackage{graphicx}
\usepackage{dcolumn}
\usepackage{bm}
\usepackage{soul}
\usepackage{color}
\usepackage{epstopdf}
\usepackage[version=3]{mhchem}
\usepackage{lipsum}
\usepackage[outercaption]{sidecap}
\usepackage{floatrow}
\usepackage{hyperref}
\usepackage{multirow}

\begin{document}


\title{Effect of the Nature of the Solid Substrate on Spatially Heterogeneous Activated Dynamics in Glass Forming Supported Films}

\author{Anh D. Phan}
\email{anh.phanduc@phenikaa-uni.edu.vn}
\affiliation{Faculty of Materials Science and Engineering, Phenikaa University, Hanoi 12116, Vietnam}
\affiliation{Phenikaa Institute for Advanced Study, Phenikaa University, Hanoi 12116, Vietnam}
\author{Kenneth S. Schweizer}
\email{kschweiz@illinois.edu}
\affiliation{Departments of Materials Science, University of Illinois, Urbana, IL 61801, USA}
\affiliation{Departments of Chemistry, University of Illinois, Urbana, IL 61801, USA}
\affiliation{Departments of Chemical $\&$ Biomolecular Engineering, University of Illinois, Urbana, IL 61801, USA}
\affiliation{Materials Research Laboratory, University of Illinois, Urbana, IL 61801, USA}
\date{\today}

\begin{abstract}
We extend the force-level ECNLE theory to treat the spatial gradients of the alpha relaxation time and glass transition temperature, and the corresponding film-averaged quantities, to the geometrically asymmetric case of finite thickness supported films with variable fluid - substrate coupling. The latter typically nonuniversally slows down motion near the solid-liquid interface as modeled via modification of the surface dynamic free energy caging constraints which are spatially transferred into the film, and which compete with the accelerated relaxation gradient induced by the vapor interface. Quantitative applications to the foundational hard sphere fluid and a polymer melt are presented. The strength of the effective fluid-substrate coupling has very large consequences on the dynamical gradients and film-averaged quantities in a film thickness and thermodynamic state dependent manner. The interference of the dynamical gradients of opposite nature emanating from the vapor and solid interfaces is determined, including the conditions for the disappearance of a bulk-like region in the film center. The relative importance of surface-induced modification of local caging versus the generic truncation of the long range collective elastic component of the activation barrier is studied. The conditions for the accuracy and failure of a simple superposition approximation for dynamical gradients in thin films is also determined. The emergence of near substrate dead layers, large gradient effects on film-averaged response functions, and a weak non-monotonic evolution of dynamic gradients in thick and cold films, are briefly discussed. The connection of our theoretical results to simulations and experiments is briefly discussed, as is extension to treat more complex glass-forming systems under nanoconfinement. 
\end{abstract}

\maketitle


\section{Introduction}
Understanding the spatially heterogeneous activated dynamics and kinetic vitrification of diverse glass-forming liquids near interfaces and under thin film confinement remains a frontier scientific challenge \cite{1,2,3,4,5,6,7,8,9,10,11,12,13,14,15,16,17,18,19,20,21,22,23,24}. Vapor and highly variable solid interfaces can induce spatial gradients of structural relaxation that can slow down or speed up relaxation, resulting in major shifts of the local and film-averaged glass transition temperature out to large length scales well beyond those that characterize changes of thermodynamics or structure near the interface \cite{1,2,3,4,5,6,7,8,9,10,11,12,13,14,15,16,17,18,19,20,21,22,23,24}. For the two-interface thin film confinement case, the surfaces can be symmetric and composed of either solids of highly variable chemistry, stiffness and topography corresponding to capped films, or two vapor interfaces corresponding to free standing films. Asymmetric films with one vapor and one solid surface are even more complex, with competing spatial gradients of dynamics of opposite nature emanating from the different interfaces. Many theoretical models have been proposed for various aspects of these systems (for reviews see refs. \cite{1,10,17}), which typically are built on very different physical ideas for the origin of glassy dynamics in the bulk. Simulations have been extensively applied to study film dynamics, with many important findings \cite{1,2,8,10,12,13,15,16,17,24}. However, computational limitations restrict simulations from approaching the ultra-long time scales probed experimentally.

Recently, a microscopic, particle level, homogeneous bulk dynamical theory has been created and widely applied to atomic, colloidal, molecular, and polymeric glass forming fluids -- the Elastically Collective Nonlinear Langevin Equation (ECNLE) theory \cite{25,26,27,28,29}. It is built on the view that the alpha relaxation process is of a coupled local-nonlocal nature, where large amplitude activated hopping on the cage scale is strongly coupled with longer range collective elastic distortions of all particles outside the cage. Based on an a priori coarse-grained mapping or complexity reduction strategy that retains key aspects of chemical specificity \cite{25,26,27,28,29}, the theory has been successfully applied to understand the temperature dependence of the alpha relaxation time of diverse families of equilibrium liquids over a range up to 14 orders of magnitude \cite{26,27,28}. Both the high temperature Arrhenius regime and the lower temperature strongly non-Arrhenius regime have been addressed in a unified and predictive manner. The central concept is the idea of a dynamic free energy that quantifies cage scale localization and the local activation barrier, and also the jump distance and emergent dynamic elastic modulus in sufficiently cold liquids that underlies the collective elastic component of the activation event.

The ECNLE theory has been extended to glass forming liquids near surfaces in thick films of both a soft (vapor) \cite{1,2,3,5,6,7,30,31,32} and hard (microscopically corrugated) nature \cite{3,4,32}. The consequences of confinement in free standing films with two vapor interfaces has been extensively worked out. Novel predictions include \cite{1,2,3,4,5,6,7,30,31,32} (i) factorization of the film location and temperature (or density) dependences of the dynamic activation barrier, (ii) strong spatial gradients of the alpha time and vitrification temperature of a double and single exponential form, respectively, (iii) a crossover to an inverse in distance from the interface power law decay of these quantities due to cutoff of the elastic field at the surface, (iv) power law decoupling of the alpha time in films from its bulk analog with apparent exponents that depend on location in the film, and (v) nonadditive gradient interference effects that emerge in thin enough films.  Essentially all these predictions have been verified by computer simulations \cite{1,2,7} for free standing thick and thin films. Experimental evidence has also been obtained for the double exponential form of the dynamical gradient close to the vapor interface of a thick film \cite{33,34}, and the results are also relevant to polymer nanocomposites \cite{35}. Predictions have been made for both microscopically rough surfaces and smooth hard walls \cite{4}.

The goal of this article is to build on these recent advances to address new aspects motivated by both their fundamental scientific interest and experimental relevance. The first new theme is to extend the theory to asymmetric films composed of one vapor interface and one solid surface. A second theme is to propose and study an effective model for the fluid-solid interface that tunes the surface dynamic free energy to mimic the highly variable and nonuniversal manner that a substrate can change the alpha relaxation process at a surface. Both aspects are studied for the foundational hard sphere fluid and a polymer melt over a wide range of film thicknesses, and most importantly for how the magnitude of the nonuniversal substrate-fluid coupling affects the dynamical gradients. How the opposite sign gradients associated with a solid and vapor surface interfere in thin films is also studied, including nonadditive collective effects beyond a linear superposition ansatz. The new model and numerical studies are broadly motivated by the experimental ability to widely tune the fluid-solid interface via topography (corrugation, smooth walls), fluid-surface potentials, and/or mechanical stiffness. Realization of the latter includes substrates composed of soft crosslinked elastomers [36], grafted polymer chains \cite{10,17,37}, and a viscoelastic liquid [38]. We suggest the new results are testable in carefully designed simulation studies which construct substrate models that specify the alpha time at the solid interface, as recently done by Simmons et. al \cite{7,24} to achieve an effectively “neutral” solid interface. Our analysis of this generalized substrate model employs the existing ECNLE theory ideas in the bulk, near interfaces, and under thin film confinement.

The article is organized as follows. In section II we first briefly review the well documented bulk and film ECNLE theories, and then present its generalization to variable substrates modeled in an implicit manner. A few model calculations are presented, along with the mapping from the reference hard sphere fluid to thermal liquids. Our core new results for the variable substrate thick and thin supported films are presented in sections III and IV, respectively. The article concludes in section V with a summary, discussion, and future outlook. All technical and conceptual details of ECNLE theory in the bulk, near surfaces, and in thin films have been documented in great detail in prior publications \cite{1,2,3,4,5,6,7,30,31,32}, and for economy of expression are not repeated. 

\section{Theory: Background and Extension to Supported Films}
\subsection{Model and dynamic free energy}
We first consider the foundational hard sphere (diameter, $d$) fluid characterized by a number density, $\rho$, and corresponding volume or packing fraction, $\Phi=\pi d^3/6$, which are taken to be the same in the bulk and under film conditions.  The basic theoretical quantity is the dynamic free energy as a function of the displacement of a tagged particle from its initial position, $r$, which is given by the sum of ideal and caging contributions which in the isotropic bulk fluid is \cite{25,26,27,28,29}
\begin{eqnarray}
F_{dyn}^{bulk}(r) &=& F_{ideal}(r)+F_{caging}^{bulk}(r),\\
\label{eq:1}
F_{ideal}(r) &=& -3k_BT\ln\left(\frac{r}{d} \right),\\
\frac{F_{caging}^{bulk}(r)}{k_BT} &=&\rho\int\frac{d\mathbf{q}}
{(2\pi)^3}\frac{S(q)C^2(q)}{1+S^{-1}(q)}\nonumber\\
\label{eq:2}
&\times&\exp\left[-\frac{q^2r^2}{6}\left(1+S^{-1}(q) \right) \right].
\label{eq:3}
\end{eqnarray}
Here, $k_B$ is Boltzmann’s constant, $T$ is temperature, $q$ is the wavevector, $C(q)=\rho^{-1}\left[1-1/S(q)\right]$ is the direct correlation function in Fourier space, and $S(q)$ is the corresponding static structure factor computed using the Ornstein-Zernike (OZ) integral equation approach with the Percus-Yevick (PY) closure \cite{39}. Within the nonlinear Langevin equation (NLE) stochastic evolution equation for a tagged particle trajectory, the quantity $-dF_{dyn}(r)/dr$  is the systematic, displacement-dependent, effective force on a moving tagged particle due to the surrounding particles \cite{25,26,27,28,29}.

Our model for supported films (thickness, $H$) with one vapor and one variable solid substrate is schematically illustrated in Fig. 1. The dynamic free energy and particle trajectories in films become a function of location in the direction ($z$) orthogonal to the interfaces and is treated as homogeneous in the two transverse directions. The minimalist theory \cite{1,2,3,5,6,7,30,31,32} employed here does not consider any equilibrium variations in spatial density or alterations of the pair structure induced by an interface corresponding to the so-called "neutral confinement" \cite{40,41,42,43,44,45,46} scenario. Hence, all surface and confinement induced changes of dynamics arise solely from kinetic considerations, not changes of structure or thermodynamics. For a single vapor or solid surface thick film ($H\rightarrow\infty$), the $z$-dependent dynamic free energy has been previously formulated in a manner that can be conceptually (not literally) described in terms of layers labeled by a discrete index $i$ with $z=(i-1)d$.  Modification of the bulk dynamic free energy is "nucleated" at the interface and transferred in a layer-by-layer or "bootstrapped" manner into the film interior as \cite{1,2,3,5,7}
\begin{eqnarray} 
F_{dyn}^{(i)}(r) &=& F_{ideal}(r) + \left(1-\frac{1}{2^i}\right)F_{caging}^{bulk}(r) + \frac{F_{caging}^{surface}(r) }{2^i}\nonumber\\
&=& F_{ideal}(r) + \frac{1}{2^i}\left(F_{caging}^{surface}(r)-F_{caging}^{bulk}(r)\right),
\label{eq:4}
\end{eqnarray}
where $F_{caging}^{surface}(r)$ is the first layer (surface) caging part of the dynamic free energy which depends strongly on the nature of the interface. Using $2^{-i}=0.5\exp\left(-\frac{z}{d}\ln{2}\right)$ reveals the underlying theoretical physical idea that caging constraints vary exponentially as a function of distance from the interface. Since the film exhibits broken symmetry, the dynamics is in principle spatially anisotropic. A full treatment of this aspect is extremely difficult. Moreover, conceptually, the dynamic local equilibrium idea underlying the construction of the dynamic free energy \cite{1,2,3,5,7} involves a minimum averaging length scale of order the cage diameter ($\sim2-3d$). Hence, in films the discrete index $i$ or distance $z$ defines the center of a cage of radius $r_{cage}$  (see Fig. 1) within which the tagged particle displacement is taken to be isotropic, corresponding to an angular averaging of motion on the cage scale.  

\begin{figure}[htp]
\includegraphics[width=8.5cm]{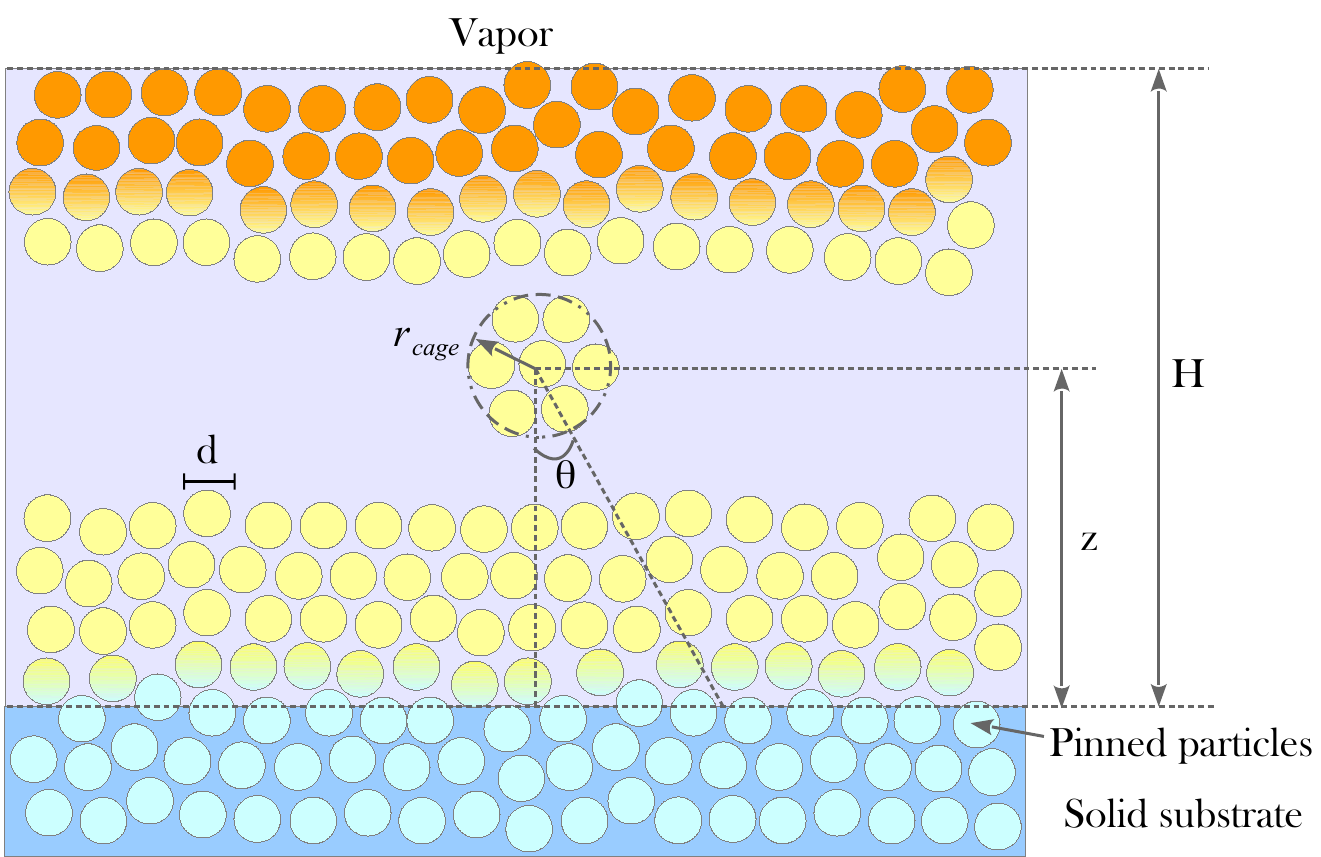}
\caption{\label{fig:1}(Color online) Schematic representation of the finite thickness supported film model and the transfer of caging constraints nucleated at the surface. For the one-surface thick film, $H\rightarrow\infty$. The cartoon of the solid interface is relevant for the “maximally rough” model of the substrate corresponding to $\kappa = 1$ where the substrate is a pinned amorphous solid configuration of the hard sphere fluid \cite{4}. The various particle colors schematically depict the variation of mobility across the polymer film.}
\end{figure}

\subsection{Influence of a rough solid substrate on dynamical caging}
The model reported in Ref. \cite{3,4} to describe vapor and solid surfaces is schematically shown in Fig.1 where the “maximally rough” substrate consists of pinned particles identical to those of the mobile liquid. Thus, half of the cage experienced by a tagged particle at the interface ($z=0$) contains pinned particles, which are modeled as randomly distributed within a full spherical cage that includes the other half of mobile particles. This is a tractability-motivated simplification that assumes the critical effect of the solid interface is the fraction of neighboring particles of a tagged particle within a cage are immobilized, rather than their precise spatial arrangement. To determine the corresponding dynamic free energy, we adopt the idea of "neutral confinement" where the pinning process does not affect the liquid pair correlation function nor induces density layering or any other change of the liquid thermodynamics. This model means that all effects of interfaces and confinement are purely dynamical, not tied to any interface-induced changes of equilibrium structure or thermodynamics.  

The solid substrate model has been generalized to an arbitrary fraction $\alpha$ of pinned particles in the surface cage. The corresponding caging part of the dynamic free energy is \cite{3,4}:
\begin{widetext}
\begin{eqnarray}
\frac{F_{caging}^{pinned}(\alpha,r)}{k_BT} &=& -\int\frac{d\mathbf{q}}{(2\pi)^3}\left[ \frac{C(q)S_{12}(q)e^{-q^2r^2/6}}{\rho(1-\alpha)\left[1- \rho(1-\alpha)C(q)\right]}+
\frac{\rho(1-\alpha)C(q)^2e^{-q^2r^2\left[2-\rho(1-\alpha)C(q) \right]/6}}{\left[1- \rho(1-\alpha)C(q)\right]\left[2- \rho(1-\alpha)C(q)\right]} \right]_{\alpha=0.5} \nonumber\\
&=& \frac{1}{2}\frac{F_{caging}^{bulk}(r)}{k_BT} + \frac{1}{2}\frac{F_{caging}^{surface}(r)}{k_BT},
\label{eq:5}
\end{eqnarray}
\end{widetext}
where $S_{12}(q)$ is collective static structure factor for density fluctuations of the immobilized and mobile particles \cite{46}:
\begin{widetext}
\begin{eqnarray}
S_{12}(q) = \frac{\rho^2(1-\alpha)\alpha C(q)}{[1-\rho(1-\alpha)C(q)][1-\rho\alpha C(q)]-\rho^2(1-\alpha)\alpha C^2(q)},
\label{eq:6}
\end{eqnarray}
\end{widetext}
Equation (5) provides a definition of the dynamic free energy associated with the solid surface. When $\alpha = 0.5$, which is our present (and prior \cite{4}) focus, one has $F_{caging}^{surface}=2F_{caging}^{pinned}(r)-F_{caging}^{bulk}(r)$, and Eq. (4) becomes
\begin{eqnarray} 
F_{dyn}^{(i)}(r) = F_{dyn}^{bulk}(r) + \frac{F_{caging}^{pined}(r) - F_{caging}^{bulk}(r)}{2^{i-1}}
\label{eq:7}
\end{eqnarray}
or in terms of the continuous variable distance from an interface as:  
\begin{eqnarray} 
F_{dyn}^{(i)}(r) = F_{dyn}^{bulk}(r) + \frac{F_{caging}^{pined}(r) - F_{caging}^{bulk}(r)}{2^{z/d}}
\label{eq:8}
\end{eqnarray}

\subsection{Model for variable substrate-fluid coupling and caging constraints}
Within the framework of ECNLE theory, we previously established [4] that a maximally rough solid substrate has a very large impact on the alpha relaxation time gradient in a thick film. A second limiting case examined was to consider a smooth hard wall where all lateral forces between the surface and fluid are absent. This limit could mimic a solid surface with a very small degree of corrugation on the length scale of fluid molecules or polymer segments, or perhaps a high surface tension liquid substrate that appears "dynamically smooth" on the alpha time scale of the confined fluid. A large reduction of dynamical constraints is expected since the surface does not exert any forces on the fluid in 2 of the 3 spatial directions. These considerations motivated our modeling of this limit by reducing the surface caging dynamic free energy by a factor of 3, i.e., $F_{caging}^{surface}=\left[2F_{caging}^{pinned}(r)-F_{caging}^{bulk}(r)\right]/3$. The theoretical predictions for this smooth hard wall model are qualitatively consistent with simulations \cite{8,15,16,40,47} which compared the dynamics of model supercooled liquids confined by a single smooth hard wall to its maximally rough substrate analog.  

The focus of the present work is to explore the predictions of ECNLE theory for films with a continuously tunable dynamical constraint at a hard surface, with the maximally rough case studied previously \cite{4}  as the limiting reference system. We do this by introducing a single parameter $\kappa$ ($0 \leq \kappa \leq 1$) to rescale the surface caging dynamic free energy to mimic different degrees of topographic surface roughness or corrugation,  or more crudely a change of substrate rigidity (e.g., an elastomer substrate or a surface with soft grafted polymers), or another modification of substrate-fluid interactions. This is an “effective”, not explicit, model of the many possible complexities at a fluid-solid interface associated with repulsive interactions. Fluid-substrate attractive interactions and adsorption are not explicitly included. However, from a highly qualitative perspective, since a major consequence of such attractions is to slow down the alpha relaxation time in the immediate vicinity of the surface, we speculate that they might be crudely mimicked by varying $\kappa$.

Mathematically, the surface caging dynamic free energy then becomes
\begin{eqnarray} 
F_{caging}^{surface}=\kappa\left[2F_{caging}^{pinned}(r)-F_{caging}^{bulk}(r)\right],
\label{eq:9}
\end{eqnarray}
where $\kappa = 1$ and 1/3 corresponds to the previously studied thick film case composed of a maximally rough and smooth solid surface, respectively \cite{4}. When $0 \leq \kappa \leq 1/3$, the caging force is even weaker than a smooth hard wall, which might mimic a lower density substrate, or systems with weaker than hard core substrate-fluid repulsive forces which could arise from microscopic mechanisms such as a rubbery surface \cite{17,36,48,49}, or an interface coated with flexible polymer grafts \cite{10,17,37}, or a viscoelastic fluid interface \cite{38}. When $\kappa=0$, there is no surface caging force corresponding to $F_{caging}^{surface}=0$, which models the previously studied vapor interface where half of the nearest neighbors of a tagged particle at the surface are missing \cite{2,3,5,6,7}. From Eqs. (\ref{eq:4}) and (\ref{eq:9}), a general expression of the gradient dynamic free energy is then
\begin{widetext}
\begin{eqnarray}
\frac{F_{dyn}(r,z)}{k_BT}&=&\frac{F_{dyn}^{bulk}(r)}{k_BT}+\frac{\kappa}{2^{z/d}}\frac{F_{caging}^{pinned}(r)}{k_BT}-\frac{1+\kappa}{2\times 2^{z/d}}\frac{F_{caging}^{bulk}(r)}{k_BT}\nonumber\\
&=& -3\ln\left(\frac{r}{d}\right)+\left(1-\frac{1+\kappa}{2}\frac{1}{2^{z/d}}\right)\frac{F_{caging}^{bulk}(r)}{k_BT}+\frac{\kappa}{2^{z/d}}\frac{F_{caging}^{pinned}(r)}{k_BT}.
\label{eq:10}
\end{eqnarray}
\end{widetext}

\begin{figure}[htp]
\includegraphics[width=8.5cm]{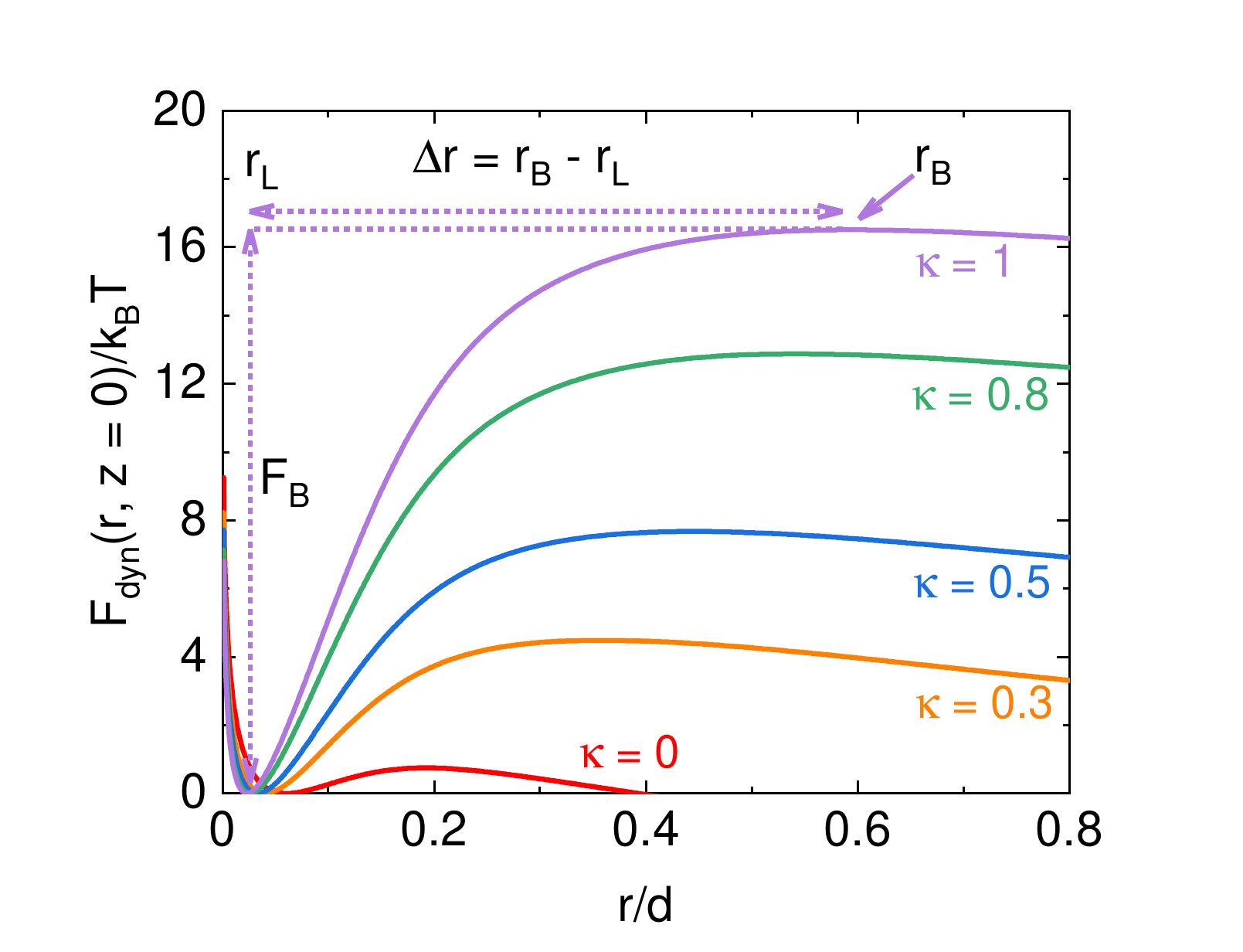}
\caption{\label{fig:2}(Color online) The dynamic free energy (in units of the thermal energy) at the substrate as a function of normalized particle displacement for a hard sphere fluid with a packing fraction $\Phi=0.57$ for different values of $\kappa$. Characteristic length and energy scales are indicated.}
\end{figure}

The dynamic free energy $F_{dyn}(r,z)$ is the crucial quantity. For bulk hard spheres, when $\Phi\geq 0.43$ a cage barrier is present which signals transient localization and a crossover to activated relaxation. Figure \ref{fig:2} shows an example of how the dynamic free energy varies with $\kappa$. Also indicated in Fig.2 are the 3 key length scales (localization length, $r_L$, barrier position, $r_B$, jump distance, $\Delta r= r_B-r_L$),  and the 3 key energy scales (harmonic curvatures $K_0=\frac{3k_BT}{r_L^2}$ and $K_B=\left|\frac{\partial^2 F_{dyn}(r,z)}{\partial r^2}\right|_{r=r_B}$  at   $r_L$ and $r_B$, respectively, and cage barrier, $F_B(z)=F_{dyn}(r_B,z)-F_{dyn}(r_L,z)$. As $\kappa$ varies from 0 to 1, the interfacial caging force increases leading to a decrease of the localization length, increase of the jump distance, and increase of the local cage barrier.  

\subsection{Collective elasticity, alpha relaxation time gradient, and mapping to thermal liquids}
The elementary alpha relaxation event in ECNLE theory is of a local-nonlocal nature where particle hopping over the cage barrier is causally coupled with a collective elastic distortion of all particles outside the cage. The physical idea is that local rearrangement requires a small cage expansion which induces a power law, scale invariant, elastic displacement field. The total barrier consists of correlated local cage and longer range collective elastic contributions. The presence of a vapor or solid interface modifies not only the dynamic free energy, but also the elastic displacement field. Here, we utilize our latest approach to this complex problem developed in Refs. \cite{2,4,5,6,7}. Diverse novel predictions based on this approach have been successfully compared with simulations of free standing films, and a supported film with a solid neutral surface \cite{2,4,5,6,7}. The adopted displacement field is taken to have the same functional form as in the bulk. It is derived using a continuum elasticity analysis modified to enforce boundary conditions whereby the amplitude vanishes at the surface (a no displacement or “zero strain” boundary condition at the interface). The result is \cite{2,4,5,6,7}
\begin{eqnarray}
u(r,\theta,z)=A_s(\theta,z)r+\frac{B_s(\theta,z)}{r^2}
\label{eq:11}
\end{eqnarray}
where $A_s(\theta,z)$ and $B_s(\theta,z)$ are chosen to enforce the boundary condition $u(r,\theta,z=0)=0$, and $\theta$ is defined in Figure \ref{fig:1}. At the cage surface where the displacement field is nucleated,
\begin{eqnarray}
u(r=r_{cage},\theta,z)&=&\Delta{r_{eff}}\nonumber\\
&=&\frac{3}{r_{cage}^3}\left(\frac{\Delta r^2r_{cage}^2}{32}-\frac{\Delta r^3r_{cage}}{192}+\frac{\Delta r^4}{3072}\right)\nonumber\\ 
&\approx&\frac{3}{32}\frac{\Delta r^2}{r_{cage}}.
\label{eq:12}
\end{eqnarray}
where $r_{cage}$ is the cage radius equal to the location of the first minimum of $g(r)$. The cage expansion amplitude is dependent on position in the film as previously derived \cite{2,4,5,6,7}:
\begin{eqnarray}
A_s(\theta,z)&=&-\frac{\Delta{r_{eff}}r_{cage}^2\cos^3\theta}{z^3-r_{cage}^3\cos^3\theta},\nonumber\\
B_s(\theta,z)&=&\frac{\Delta{r_{eff}}(z)r_{cage}^2z^3}{z^3-r_{cage}^3\cos^3\theta},
\label{eq:13}
\end{eqnarray}
The corresponding elastic barrier follows by integrating over all film particles outside the cage (the center of which is at a distance $z$ from the surface)  \cite{2,4,5,6,7}: 
\begin{eqnarray}
F_{e}=\frac{3\Phi}{\pi d^3}\int_{V_{film}}d\textbf{r}u^2(r,\theta,z)K_0(r,z).
\label{eq:14}
\end{eqnarray}

Before proceeding, we briefly review the physical ideas \cite{1,2,3,4,5,6,7,30,31,32} that underlie our approximate modeling of the elastic displacement field and calculation of the elastic barrier for vapor and solid interfaces. First, we emphasize that using continuum mechanics to construct the functional form of the elastic displacement field in our molecular-based theory is a simplification. A continuum perspective is not used to quantify the amplitude of the displacement field, nor compute the magnitude of the elastic barrier. Rather, the latter are done using the microscopic dynamic free energy idea, and the Einstein glass picture of localized particles outside the cage \cite{29}. Such a "mixed" microscopic-continuum analysis of the collective elastic barrier contrasts with phenomenological elastic models in the literature, e.g., the "shoving model" \cite{25,26,27,28,29}. Concerning boundary conditions, we have adopted the vanishing of the elastic displacement field at the interface (akin to a “zero strain” condition in mechanics) for both vapor and solid interfaces. This is a natural and simple choice for a rigid solid surface, or perhaps if there are strong surface-liquid attractions at the solid interface although we are not explicitly considering this case. 

An alternative possible choice for a vapor interface is a “zero stress” boundary condition. We have not done this, with our motivation as follows. (i) Since the displacement fields are of low amplitude, a linear elastic scenario applies where stress and strain are proportional, so naively the adopted zero-displacement condition might be crudely viewed as in the same spirit as it zero-stress analog. (ii) If surface tension plays a dominant dynamical role at a vapor interface, the zero-displacement boundary condition may be suitable since it minimizes surface area under elastic deformation. (iii) Given the “mixed” continuum-microscopic nature of our analysis of the elastic barrier, a priori whether a zero-strain or zero-stress boundary condition is more appropriate is unclear to us. Of course, none of the above arguments are rigorous. But we emphasize that use of the zero-displacement boundary condition for free standing films has led to many striking predictions by ECNLE theory for the role of collective elasticity in thick and thin films, which have major support from simulations (Refs. \cite{1,2,3,6,7}). At the most fundamental level, one can view the problem as open and requiring more work, and whether for vapor surfaces the zero-stress boundary condition may be more appropriate could depend on the questions asked within the ECNLE theory framework. 

Now, by combining all the above results, the alpha or structural relaxation time gradient is determined using the Kramer’s mean first passage time for barrier crossing expression \cite{2,4,5,6,7} 
\begin{eqnarray}
\frac{\tau_\alpha(z)}{\tau_s} = 1+ \frac{2\pi}{\sqrt{K_0(z)K_B(z)}}\frac{k_BT}{d^2}\exp\left(\frac{F_B(z)+F_e(z)}{k_BT}\right),\nonumber\\
\label{eq:15}
\end{eqnarray}
where $\tau_s$ is a "short time/length scale" non-activated dissipative process taken to be unaffected from interfaces and confinement, and is given by \cite{2,4,5,6,7,26}:
\begin{eqnarray}
\tau_s &=& \tau_0g(d)\left[1+\frac{1}{36\Phi}\int_{0}^{\infty}dq \frac{q^2\left[ S(q)-1 \right]^2}{S(q)+b(q)} \right], \nonumber\\
\tau_0&=& \frac{1}{24\rho d^2}\sqrt{\frac{M}{\pi k_BT}},
\label{eq:16}
\end{eqnarray}
where $b^{-1}(q) = 1-j_0(q)+2j_2(q)$, $j_n(x)$ is the spherical Bessel function of order $n$, $M$ is the particle mass, and $\tau_0$ is a "bare" Boltzmann-like time scale. 

To quantitatively apply ECNLE theory for hard sphere fluids to experiments and simulations on thermal liquids we employ the well-developed coarse-grained mapping or complex reduction strategy to a thermalized effective hard sphere model. The physical idea is to equate the dimensionless compressibility (rigorously equal to the non-dimensionalized amplitude of long wavelength thermal density fluctuations) predicted by integral equation theory for hard-sphere fluids, $S_0^{HS}$, and the corresponding experimental quantity, $S_0(T)=\rho k_BT\kappa_T$, deduced from equation of state data for liquids in equilibrium. A key physical idea for molecules and polymers is that it is the amplitude of density fluctuations at the level of the number of elementary rigidly moving interaction sites, $N_s$, that enters the mapping. The resulting mapping relation based on the OZ-PY approach is \cite{2,4,5,6,7,26,27,28}: 
\begin{eqnarray}
S_0^{HS}=\frac{\left(1-\Phi_{eff}\right)^4}{\left(1+2\Phi_{eff}\right)^2}\equiv S_{0,expt}\approx \frac{1}{N_s}\left(-A+\frac{B}{T}\right)^{-2}
\label{eq:17}
\end{eqnarray}
Eq. (\ref{eq:17}) determines $\Phi_{eff}$ as a temperature and chemistry-specific effective hard sphere fluid packing fraction. The second approximate equality has been shown to be highly accurate \cite{26,27,28}, with $A$ and $B$ corresponding to an interaction site level defined entropic packing and cohesive energy quantities, respectively. As mentioned above, $N_s$ is the number of interaction sites for a rigid molecule or the corresponding number in a Kuhn segment of a polymer chain.  Using this mapping, the mean alpha time can be predicted with no adjustable parameters. In this article, we present results both for the literal hard sphere model, and for the widely studied thermal polystyrene (PS) melt for which prior work \cite{2,4,5,6,7,28} has shown $d=1.16$ nm (Kuhn segment space filling diameter), A = 0.618, $B = 1297$ $K$, and $N_s=38.4$.
\subsection{Confined thin films}
We model a finite thickness supported film with a vapor surface at $z=H$ as illustrated in Fig. \ref{fig:1}. The same basic analysis for a single interface thick film can be employed to describe a vapor layer. We have proposed that the combined effects of the two interfaces on the dynamic free energy can be superimposed at the fundamental level of the dynamic free energy \cite{2}. Importantly, this does not mean that physical quantities predicted by the theory (e.g., alpha time, $T_g$ gradients) obey naive superposition. Given this idea, Eq. (\ref{eq:10}) for the dynamic free energy of the tagged particle in the $i^{th}$ layer can then be written as	
\begin{widetext}
\begin{eqnarray}
\frac{F_{dyn}(r,z,H)}{k_BT}= -3\ln\left(\frac{r}{d}\right)+\left(1-\frac{1+\kappa}{2}\frac{1}{2^{z/d}}\right)\frac{F_{caging}^{bulk}(r)}{k_BT}+\frac{\kappa}{2^{z/d}}\frac{F_{caging}^{pinned}(r)}{k_BT}-\frac{1}{2\times2^{(H-z)/d}}\frac{F_{caging}^{bulk}(r)}{k_BT}.
\label{eq:18}
\end{eqnarray}
\end{widetext}
As before, the local and elastic barriers can be calculated as a function of position in the film, the overall film thickness, and temperature to predict the gradient of the alpha relaxation times and glass transition temperatures of finite-size supported films.
\section{Thick Supported Films}
\subsection{Analytic analysis}
Before presenting our numerical results for thick supported films, we first employ the "ultra-local" analytic analysis method \cite{50} developed for bulk hard sphere fluids to construct an initial qualitative understanding of the effect of $\kappa$ on the most elementary feature in films: the dynamic localization length. This analysis also provides insights concerning the local cage barrier \cite{46,50}. 

The ultra-local analytic analysis is based on the approximate dominance of high wavevector contributions ($q>q_c$) to the dynamic force vertex in Eq. (\ref{eq:10}) for systems with a sufficiently high cage barrier. Specifically, in the high wavevector regime, $S(q)\approx 1$ and based on the PY closure for hard sphere fluids \cite{39}, $C(q)\approx-4\pi d^3\cos(qd)(qd)^{-2}$. Using these simplifications in the condition that defines the minimum of the dynamic free energy and hence localization length, $\left.\frac{\partial F_{dyn}(r,z)}{\partial r}\right|_{r=r_L}=0$, one can easily derive an expression for the inverse dynamic localization length: 
\begin{widetext}
\begin{eqnarray}
\frac{3d}{r_L}=\frac{8\Phi g(d)^2r_L}{\pi}\left[ \int_{q_c}^{\infty} dq e^{-q^2r_L^2/3}\left(1-\frac{1+\kappa}{2}\frac{1}{2^{z/d}} \right) + \int_{q_c}^{\infty} dq e^{-q^2r_L^2/6}\left(\frac{1}{2}+\frac{1}{2}e^{-q^2r_L^2/6} \right)\frac{\kappa}{2^{z/d}} \right].
\label{eq:19}
\end{eqnarray}
\end{widetext}

Since localization lengths must be small compared to the inverse crossover wavevector for the applicability of the ultra-local analysis \cite{46,50}, one has $q_cr_L\ll 1$, and hence the lower limit of the integral of Eq. (\ref{eq:19}) can be approximated as zero. Equation (\ref{eq:19}) then simplifies to:
\begin{eqnarray}
\frac{3d}{r_L}=\frac{4\sqrt{3}\Phi g(d)^2}{\sqrt{\pi}}\left[ \left(1-\frac{1+\kappa}{2}\frac{1}{2^{z/d}}\right)+ \left(\frac{\sqrt{2}}{2}+\frac{1}{2} \right)\frac{\kappa}{2^{z/d}}\right].\nonumber\\
\label{eq:20}
\end{eqnarray}
From this, the dynamic localization length follows as
\begin{eqnarray}
r_L(z)=\frac{r_{L,bulk}}{1-\cfrac{1-\kappa}{2}\cfrac{1}{2^{z/d}}+\cfrac{\sqrt{2}-1}{2}\cfrac{\kappa}{2^{z/d}}}.
\label{eq:21}
\end{eqnarray}
Equation (\ref{eq:21}) predicts that $1/r_L(z)$  scales linearly with $\kappa$, i.e., stronger substrate caging constraints lead to smaller dynamic localization lengths, a physically expected trend. For the $\kappa=0$ vapor surface one has, $r_L(z)=\cfrac{r_{L,bulk}}{1-\cfrac{1}{2}\cfrac{1}{2^{z/d}}}$ and $r_L(z)=2r_{L,bulk}$. When $\kappa=1$, the surface is a maximally rough solid and $r_L(z)=\cfrac{r_{L,bulk}}{1+\cfrac{\sqrt{2}-1}{2}\cfrac{1}{2^{z/d}}}$, which at the interface ($z=0$), becomes $r_L(z)=\cfrac{2r_{L,bulk}}{1+\sqrt{2}}$. Interestingly, for $\kappa=1/\sqrt{2}\approx0.707$, we find $r_L(z)=r_{L,bulk}$, and hence the localization length (and the corresponding localization well harmonic spring constant $K_0(z)$) equal their bulk values unaffected by the solid surface at all film locations. This striking prediction is discussed further below. More generally, the alpha time and vitrification temperature gradients require $K_0(z)$, $K_B(z)$, $F_B(z)$, and $\Delta r(z)$ (see Fig. \ref{fig:2}) which are determined numerically from the theoretically predicted $F_{dyn}(r,z)$ in Eq.(\ref{eq:10}).
\subsection{Local cage and collective elastic barrier}
We first illustrate the role of $\kappa$ on dynamical properties for a hard sphere fluid packing fraction for which the homogeneous bulk fluid alpha time is approximately 100 ns. This choice is motivated by the fact that solid surfaces typically strongly increase the relaxation time, and also by the fact that 100 ns is a typical longest relaxation time probed in standard molecular dynamics simulations. This timescale is achieved based on bulk ECNLE theory \cite{7,25,26} for a packing fraction of 0.5788.

\begin{figure}[htp]
\includegraphics[width=8.5cm]{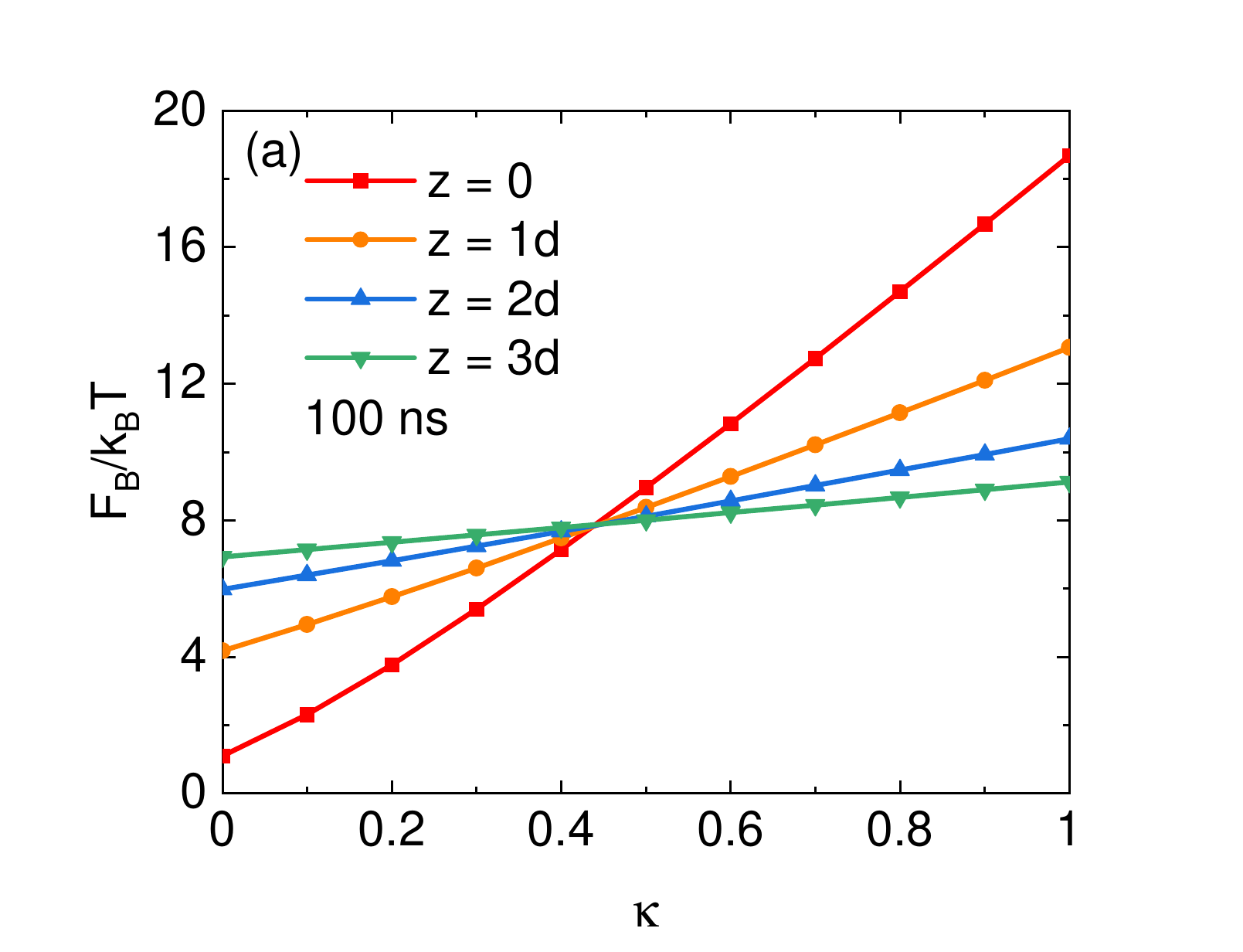}\\
\includegraphics[width=8.5cm]{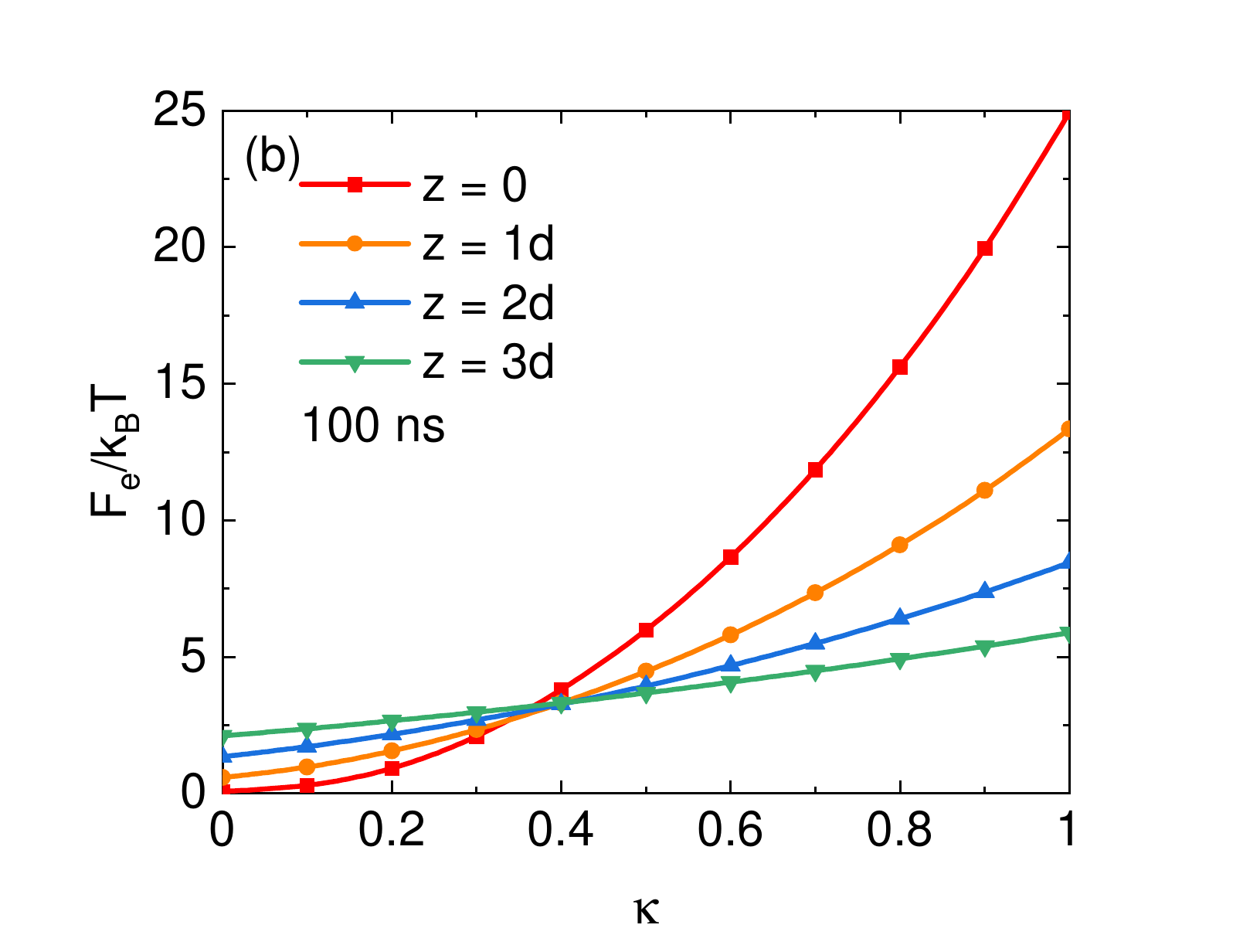}
\caption{\label{fig:3}(Color online) (a) The local cage barrier and (b) collective elastic barrier as a function of $\kappa$ for $\Phi=0.5788$  corresponding to a bulk alpha time of ~100 ns, at different positions in the thick film. The bulk local cage and bulk elastic barriers are $F_B^{bulk}\approx 7.9k_BT$ and $F_e^{bulk}\approx 3.87k_BT$, respectively. }
\end{figure}

Figure \ref{fig:3}a shows the local cage barrier over the entire range of $\kappa$ of present interest, at four locations near the surface. The cage barrier grows with $\kappa$ in a weakly nonlinear manner, with a slope that decreases with distance from the interface. These trends are expected, however the near intersection of all the curves at $\kappa\approx0.44$ is surprising. The cage barrier at the near intersection is close to its bulk value of $7.9k_BT$, suggesting a subtle “cancellation” of surface effects. Further insight to this and other behaviors in Fig. \ref{fig:3}a can be gleaned from our analytic analysis.

First recall that based on using the OZ-PY theory for the fluid structure as input to the ECNLE theory, a near linear proportionality between $F_B$ and $1/r_L$ was previously discovered and analytically understood \cite{46,50}. Operationally, our numerical results in Fig. \ref{fig:3}a are roughly consistent with a linear relation $F_B(z)\sim\kappa$,  as expected if the cage barrier scales as the inverse localization length given in Eq. (\ref{eq:21}). This $F_B-\kappa$ linearity also strongly suggests that the dependence of $r_B$ and $r_L$ on $\kappa$ plays a minor role in the determination of $F_B$. The slope of the lines in Fig. \ref{fig:3}a significantly decreases deeper into the film, as expected from Eq. (\ref{eq:21}) and the linear $F_B(z)\sim\kappa$ connection. 

The insensitivity of $F_B(z)$ to position in the film for $\kappa\approx0.44$ can be qualitatively explained mathematically using Eq. (\ref{eq:10}). This equation can be rewritten as
\begin{widetext}
\begin{eqnarray}
F_{dyn}(r,z)=F_{dyn}^{bulk}(r)+\frac{2F_{caging}^{pinned}(r)-F_{caging}^{bulk}(r)}{2\times2^{z/d}}\left[\kappa-\frac{F_{caging}^{bulk}(r)}{2F_{caging}^{pinned}(r)-F_{caging}^{bulk}(r)} \right].
\label{eq:22}
\end{eqnarray}
\end{widetext}
If $F_{dyn}(r,z)$ is nearly independent of position near the solid surface, then $\kappa$ is
\begin{eqnarray}
\kappa=\frac{F_{caging}^{bulk}(r)}{2F_{caging}^{pinned}(r)-F_{caging}^{bulk}(r)}.
\label{eq:23}
\end{eqnarray}
In the ultra-local analytic limit, we have
\begin{eqnarray}
\frac{F_{dyn}^{bulk}(r)}{k_BT}&=&-\frac{12\Phi g^2(d)}{\pi d}\int_{q_c}^{\infty}\frac{dq}{q^2}e^{-q^2r^2/3},\\
\label{eq:24}
\frac{F_{dyn}^{pinned}(r)}{k_BT}&=&-\frac{24\Phi g^2(d)}{\pi d}\int_{q_c}^{\infty}\frac{dq}{q^2}\left[\frac{e^{-q^2r^2/6}}{2}+\frac{e^{-q^2r^2/3}}{4}\right],\nonumber\\
\label{eq:25}
\end{eqnarray}
Substituting Eqs. (\ref{eq:24}) and (\ref{eq:25}) into Eq. (\ref{eq:23}), and using $q_cr\ll 1$, then yields $\kappa=\cfrac{1}{2e^{q_c^2r^2/6}}$. Since the barrier position $r_B$ falls in the particle displacement range of $0.4d$ to $0.6d$ when $\Phi=0.5-0.61$ (Fig. \ref{fig:2}), the inequality $q_cr\ll 1$ is roughly applicable (though not overly strong). Based on these arguments, tuning $\kappa$ to a value just below 0.5 results in $F_{dyn}(r,z) \rightarrow F_{dyn}^{bulk}(r)$. Our full calculations in Fig. \ref{fig:3}a suggest that the dynamic gradient vanishes for $\kappa\approx0.44$.  Of course, quantitative deviations between the full numerical calculations and the analytic analysis are expected since the latter is approximate.

At a more detailed level, one can ask how the value of $\kappa$ at the intersection point varies with packing fraction. In prior work \cite{46}, the characteristic wavelength was approximated as $q_c=\cfrac{2\pi}{r_{cage}}$. Increasing packing fraction weakly decreases the cage radius $r_{cage}$, thereby slightly increasing $q_c$, leading to a weak decrease of $\kappa$ at the intersection point. But overall, this is a small effect. 

The elastic barrier depends not only on the finite-size nature of the films via the geometric cutoff of the elastic field at an interface, but also on the nature of the surfaces via the dynamic free energy properties which enter the elastic barrier (harmonic spring constant, jump distance) in a $z$-dependent manner \cite{2,4,5,6,7}. To compare with $F_B(z)$ in Fig. \ref{fig:3}a, Figure \ref{fig:3}b shows the dependence of $F_e(z)$ on $\kappa$ near the solid surface. In contrast to the cage barrier, the elastic barrier does not grow linearly with $\kappa$ since its magnitude is related to the jump distance and harmonic curvature, not the inverse localization length. However, interestingly, $F_e(z)$ is roughly independent of $z$ when $\kappa\approx0.4$. This finding is qualitatively consistent with a relatively uniform ($z$-independent) local cage barrier in the first few layers near the surface when $\kappa\approx0.44$, per Fig. \ref{fig:3}a. The differences in this behavior of the two barriers are attributed to a greater influence of interfacial effects on the elastic barrier compared to its local barrier analog. Overall, the different dependences of the local and elastic barriers on $\kappa$ imply the theory predicts tuning this substrate parameter will modify the relative importance of local caging and long-range collective elasticity, which will have multiple dynamical consequences. For example, how interfaces modify dynamic fragility, and the quantitative aspects of the alpha time and glass transition temperature spatial gradients.

\subsection{Alpha time and glass transition temperature gradients}
Previously we showed \cite{7} that ECNLE theory predicts the alpha time gradient is of a double exponential decay form near the interface, which then crosses over to an inverse power law decay far enough from the interface before attaining its bulk fluid limit. Moreover, the glass transition temperature displays an exponential dependence on $z$, also with a weak power law tail.  Within the present formulation of ECNLE theory for symmetry-broken films, these core predictions are qualitatively identical for all surface boundary conditions, and have been verified in simulations \cite{1,2,7,24,40,41,42,43,44,45,51} and experiment \cite{1,33,34,35}. Hence, in this section on thick films, we first we focus on the dependence of the alpha time and $T_g$ on $\kappa$ only at selected locations in the thick film.

\begin{figure}[htp]
\includegraphics[width=8.5cm]{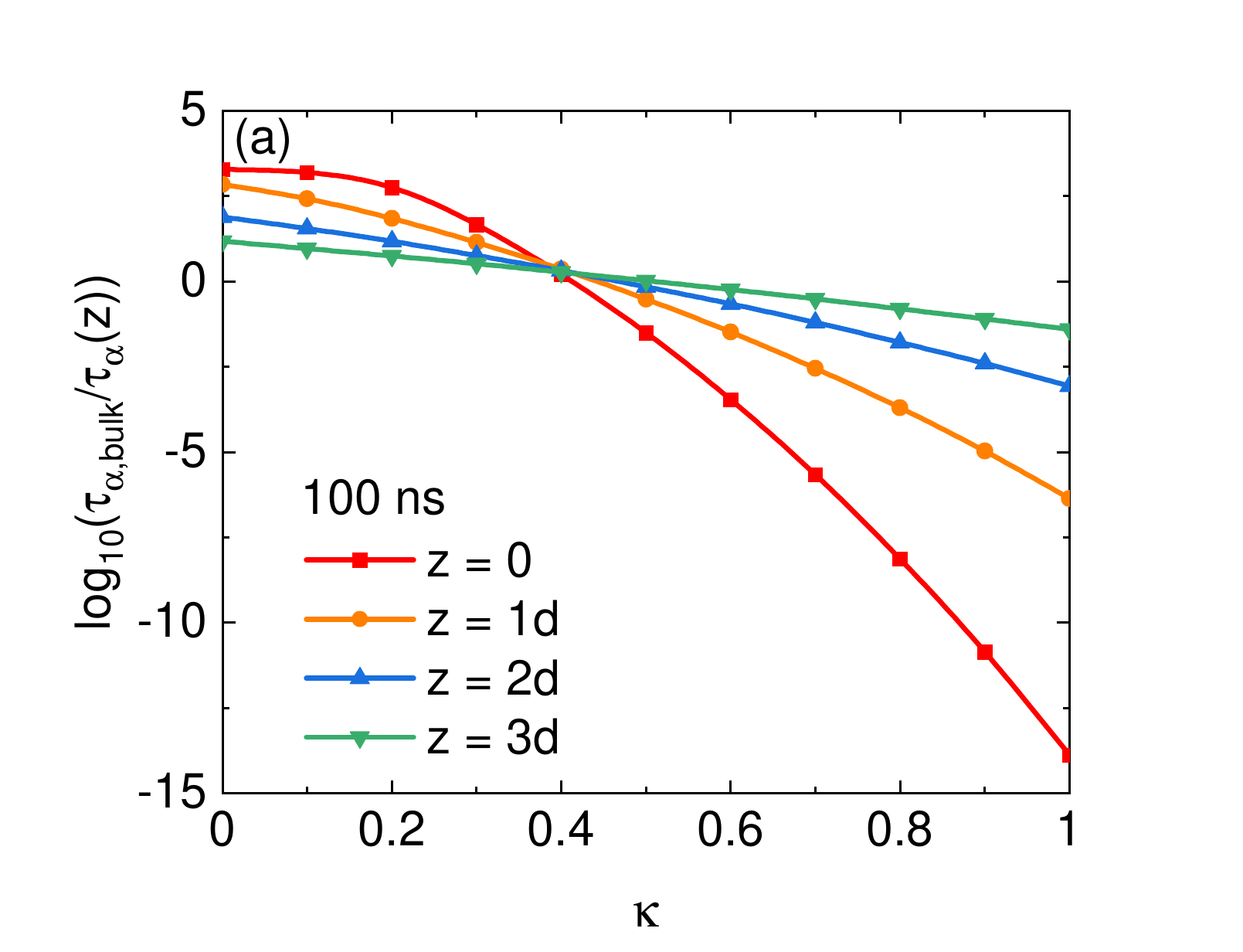}\\
\includegraphics[width=8.5cm]{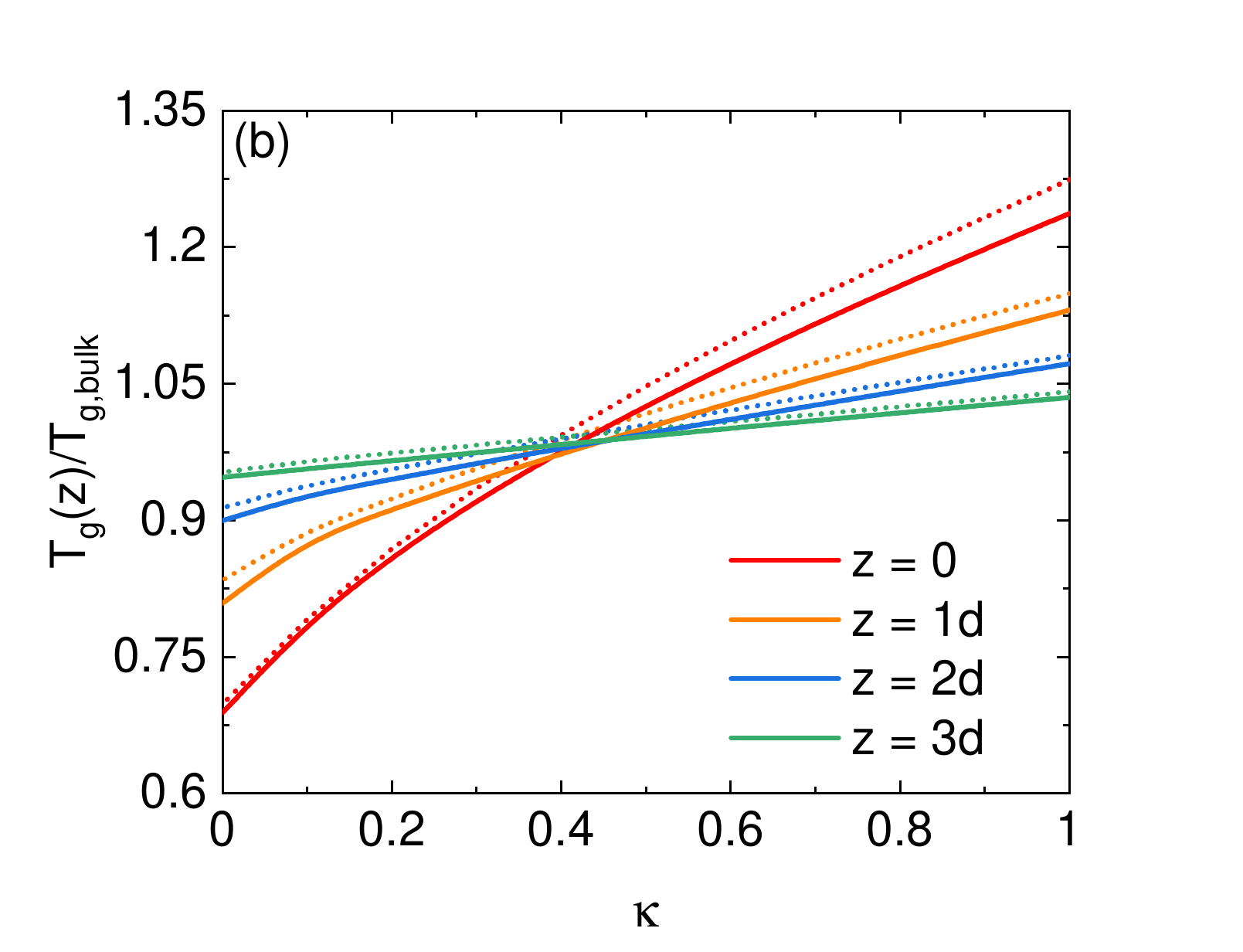}
\caption{\label{fig:4}(Color online) (a) Logarithm of the inverse alpha time, and (b) local glass transition temperature, of polystyrene melts, normalized by their bulk counterparts, near the substrate surface as a function of $\kappa$. The solid and dotted curves correspond to calculations based on vitrification timescale criteria of 100 s and 100 ns, respectively. }
\end{figure}

\begin{figure}[htp]
\includegraphics[width=8.5cm]{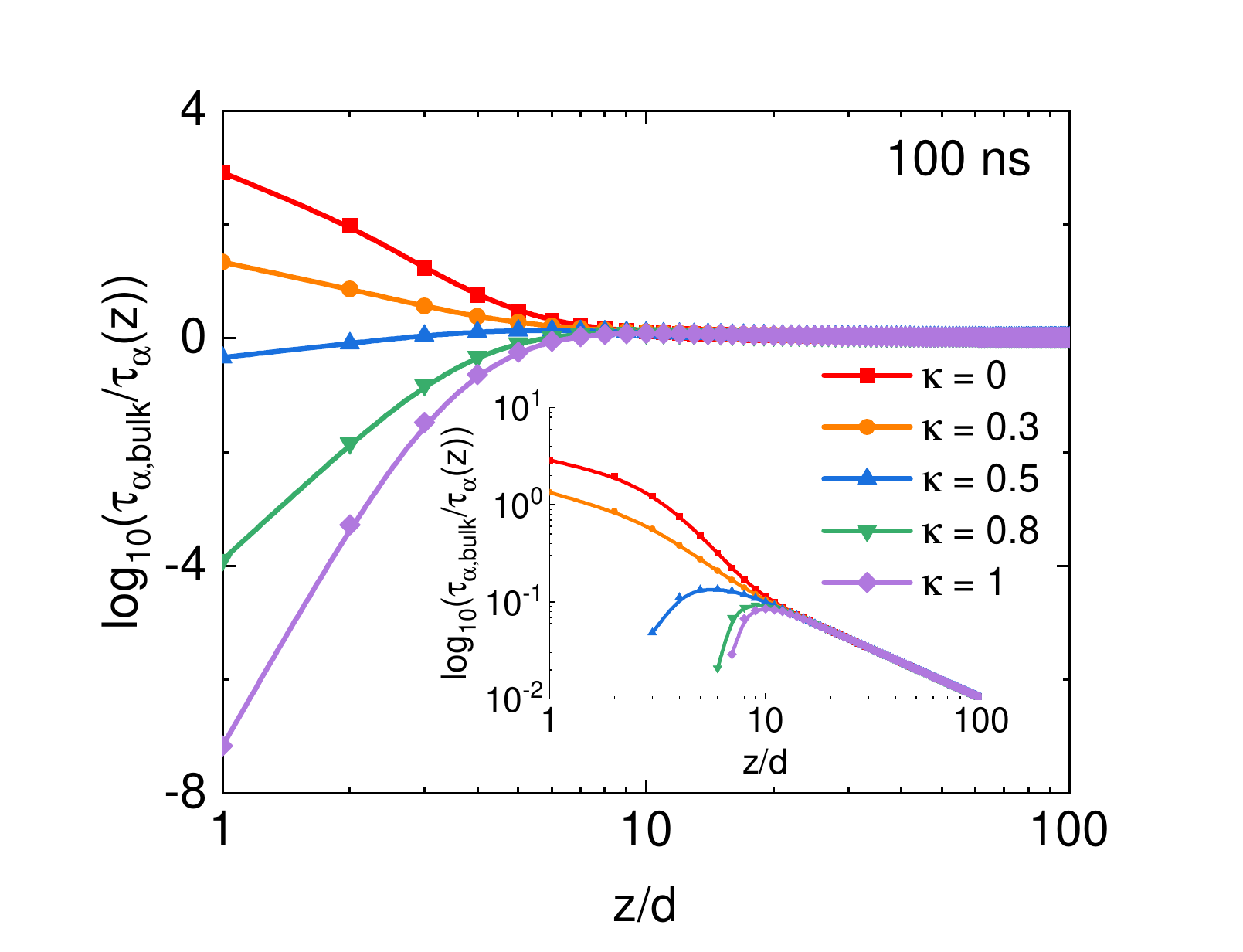}
\caption{\label{fig:5}(Color online) Logarithm of the inverse alpha time normalized by its bulk value for PS films at $T_{g,bulk}\approx502.6$ $K$ (or $\Phi=0.5788$) based on the indicated 100 ns kinetic vitrification criterion, as a function of distance from the solid surface, with $\kappa = 0, 0.3, 0.5, 0.8,$ and 1.0. Inset: identical results as the main frame but presented in a double-logarithmic format.}
\end{figure}

Figure \ref{fig:4} shows the normalized alpha relaxation time and glass transition temperature near the solid interface of a thick film. The local $T_g$ is determined based on two dynamic criteria that mimic experimental and simulation studies, $\tau_\alpha(T_g(z))=100$ s and 100 ns, respectively. As expected, increasing $\kappa$ induces more slowing down (Fig. \ref{fig:4}a), and in a manner that varies in a supra-exponential manner with $\kappa$. This behavior results in the glass transition temperature gradient trends in Fig. \ref{fig:5}b, and generally a nonlinear variation with $\kappa$.  Interestingly, we again find an "invariant" to interface state at $\kappa\sim0.42$, qualitatively consistent with results in Fig. \ref{fig:3}a and \ref{fig:3}b. 

In prior works, it was proved that within ECNLE theory the $T_g$ gradient normalized by its bulk value in free-standing films, and also in supported films with rough or smooth solid surfaces, is nearly unaffected by the vitrification criterion over a wide range of timescales \cite{4,5,6}. Thus, we expect such an insensitivity of $T_g(z)/T_{g,bulk}$ holds as $\kappa$ is varied, which is confirmed in Fig. \ref{fig:4}. Note that the $T_g(z)/T_{g,bulk}$ curves corresponding to $\tau_\alpha(T_{g,bulk})=100$ s and 100 ns determined as a function of $\kappa$ near the solid interface are close. This suggests that the normalized $T_g$ gradients determined from simulations can be utilized to predict the gradient under typical experimental conditions. 

Figure \ref{fig:5} shows calculations of the spatial gradient of the inverse alpha time normalized by its bulk counterpart as a function of distance from the substrate for a PS thick film at a temperature of $T_{g,bulk}$. The results in the main frame are presented in a log-linear format. Near the interface ($z\leq0.5d$), the normalized mobility gradient obeys a double-exponential spatial form for all surfaces. For $\kappa-0.5$, one sees in the inset of Fig.\ref{fig:5} that $\log_{10}\left(\tau_{\alpha,bulk}/\tau_{\alpha}(z)\right)\approx0.05-0.15$ over a large range of film locations spanning from $z \sim 0$ to $8d$. The range of variation of this normalized mobility gradient is "flattened" when $\kappa=0.4$, consistent with the alpha time being nearly $z$-independent (Fig. \ref{fig:4}a). Far from the interface ($z \geq 10d$), the double-exponential behavior is replaced by an inverse power law form that was previously predicted and observed in simulation for a thick free-standing film \cite{7}. The existence of this inverse power law contribution is why a double exponential form for the alpha time gradient in thick films and its linear superposition cannot describe numerous dynamic features at the center of thin films, as discussed in Ref. \cite{2}. Since a solid and vapor interface truncate the collective elastic field in the same manner within the present theory, the power-law tail decay is identical in functional form for different interfaces. 

Concerning the inset of Fig. \ref{fig:5}, there is a striking degree of quantitative universality of the power law tail. This is understandable since this long-range tail is determined by the universal cutoff of the elastic field at an interface mechansim, and that far enough from the interface the dynamic free energy attains its bulk liquid form. To elaborate, recall that we previously derived far from an interface that $\log_{10}\left(\cfrac{\tau_{\alpha,bulk}(T)}{\tau_{\alpha}(T,z)}\right)\approx\log_{10}e\cfrac{F_{e,bulk}}{k_BT}\cfrac{r_{cage}}{4z}$ \cite{7}. The tail amplitude is not universal with regards to chemistry since it depends on the elastic barrier in the bulk, the magnitude of which plays an essential role in determining the dynamic fragility in ECNLE theory \cite{4,5,29}. However, in the present context, it is indeed independent of $\kappa$ which nucleates a change of the dynamic free energy at the interface that is transferred into the film interior, but it effect decays exponentially as stated above. Thus, the magnitude of $\kappa$ does strongly modify both the local cage and collective elastic barriers close enough to the interface where the dynamic free energy is spatially heterogeneous, but far enough from the interface where the power law form of the gradient cleanly emerges due to the pure cutoff effect the dependence on $\kappa$ vanishes, per the good overlap of curves region in the inset of Fig. \ref{fig:5}.  

The above findings suggest a simple analytic (though approximate) form for the alpha time gradient discussed previously \cite{7}:
\begin{eqnarray}
\log_{10}\left(\cfrac{\tau_{\alpha,bulk}(T)}{\tau_{\alpha}(T,z)}\right) = A_\kappa(T)e^{-z/\xi_\kappa(T)} + \frac{B_\kappa(T)}{z}.
\label{eq:26}
\end{eqnarray}
Here, $A_\kappa(T)$ is the amplitude at the surface, $\xi_\kappa(T)$ is the penetration or decay length (predicted by ECNLE theory to be nearly independent of temperature \cite{4,5}), and $B_\kappa(T)$ is the amplitude of the universal in form power law decay part of the gradient due solely to the interfacial cut-off of the long-ranged collective elastic field. We shall use this expression in section IV to study deviations from the naive gradient superposition approximation. 

Concerning the accuracy of Eq. (\ref{eq:26}) within the ECNLE theory framework, we note that prior work \cite{7} on thick supported films with one neutral solid surface and one vapor interface has verified Eq. (\ref{eq:26}) provides a good representation of the numerical ECNLE theory predictions. In addition, based on the present formulation of collective elasticity effects in ECNLE theory for symmetry-broken films, we analytically know that very far from an interface in thick films that  $\log_{10}\left(\cfrac{\tau_{\alpha,bulk}(T)}{\tau_{\alpha}(T,z)}\right)\approx\log_{10}e\cfrac{F_{e,bulk}}{k_BT}\cfrac{r_{cage}}{4z}$ \cite{7}. This result implies that $B_\kappa=\log_{10}e\cfrac{F_{e,bulk}}{k_BT}\cfrac{r_{cage}}{4}$ in Eq. (\ref{eq:26}) is always positive. Hence, as the bulk behavior is approached far from the interface, the dynamics is slightly faster than in the bulk. On the other hand, $A_\kappa$ can be positive or negative, and strongly depends on a nature of the interface; a negative $A_\kappa$ is the dominant origin of slowing down of relaxation near a solid interface. Taking Eq. (\ref{eq:26}) seriously would then appear to suggest that a liquid near a solid surface could exhibit a different glass transition behavior compared to the behavior near the interface or in the film center. As a speculative comment, such opposite effects of $A_\kappa$ and $B_\kappa$ on glassy dynamics could be a possible origin of the observation of two $T_g$'s in the pore confined liquids studied by McKenna and coworkers \cite{52,53}. More generally, one expects within ECNLE theory that the suppression of dynamics near solid surfaces becomes more significant if particles are confined within (spherical-like, or cylindrical) pores since such a change of geometry amplifies the coupled dynamical effects of interfacial cut-off of the long-ranged collective elastic field and modification of the dynamic free energy.  

\section{Confined Thin Films}
Within ECNLE theory, confinement modifies the structural relaxation time and glass transition temperature via two distinct physical effects \cite{2}: (i) strengthening/weakening of the local caging constraint near interfaces which is embedded in the surface dynamic free energy and spatially transferred into the film, and (ii) a reduction of the collective elasticity cost for cage scale hopping due to the interfacial cutoff of the elastic displacement field.  Here we explore the evolution of these effects for supported films of variables thicknesses with different interfaces as a function of $\kappa$. 
\subsection{Glass transition temperature gradients}
\begin{figure}[htp]
\includegraphics[width=8.5cm]{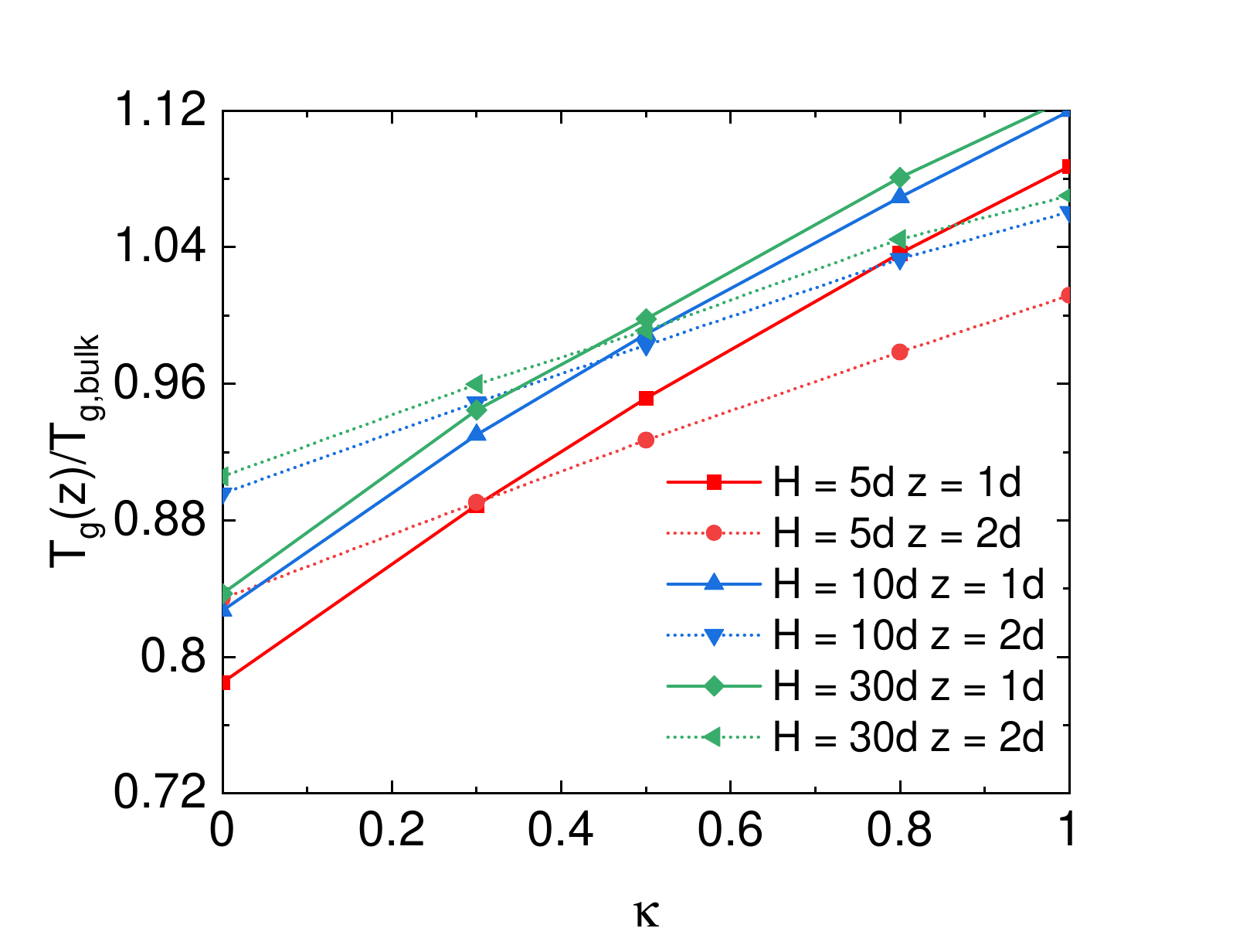}
\caption{\label{fig:6}(Color online) The local glass transition temperatures of PS films of thicknesses $H = 5d, 10d$, and $30d$, normalized by their respective bulk values, as a function of $\kappa$. The solid and dotted curves correspond to calculations at $z = 1d$ and $2d$, respectively. The bulk $T_g$ equals 428.3 $K$ and corresponds to the typical experimental vitrification time scale of 100 s.}
\end{figure}

We first calculate the spatially resolved ratio of the vitrification temperature $T_g(z)/T_{g,bulk}$ as a function of $\kappa$ near the substrate for various film thicknesses. Since we have previously shown that the $T_g$ gradient normalized by its bulk value remains nearly unchanged with modification of vitrification criterion, we adopt the experimental vitrification time scale of 100 s. For the PS liquids of present interest, the latter criterion corresponds to $T_{g,bulk}=428.3$ $K$. The corresponding $T_g(z)$ curves for polymer films are shown in Fig. \ref{fig:6} for $H=10d$ and $30d$ at the near surface locations of $z = 1d$ and $z = 2d$. Note that the curves intersect at $\kappa=0.42$, a value close to what was found for $T_g(z)$ in thick supported polymer films (Fig. \ref{fig:4}b). This finding implies that near a solid substrate with $\kappa=0.42$, there is an interfacial region with a nearly uniform distribution of glass transition temperatures. However, as the film becomes thinner, $T_g(z)$ decreases due to the cutoff of the elastic barrier at the surfaces. For the polymer film with $H=5d$, $T_g(z)/T_{g,bulk}\sim0.9$ in the first few interfacial layers for $\kappa = 0.3$ since both the local and elastic barrier are significantly reduced. 

The nearly linear relationship in Fig. \ref{fig:6} between $T_g(z)/T_{g,bulk}$ and $\kappa$ at $z=2d$ can be understood based on Eq. (15) which indicates how $T_g(z)/T_{g,bulk}$ depends on variations of the local and elastic barriers as a function of $\kappa$. As seen in Fig. \ref{fig:3}, both $F_B$ and $F_e$ are proportional (to leading order) to $\kappa$ beyond the first two layers from the solid interface. On the other hand, very close to the surface at $z=0$ or $1d$, the gradient $T_g(z)/T_{g,bulk}$ does not vary linearly with $\kappa$. The origin of this nonlinearity is attributed to the strongly nonlinear relationship between $F_e/k_BT$ and $\kappa$.

\begin{figure}[htp]
\includegraphics[width=8.5cm]{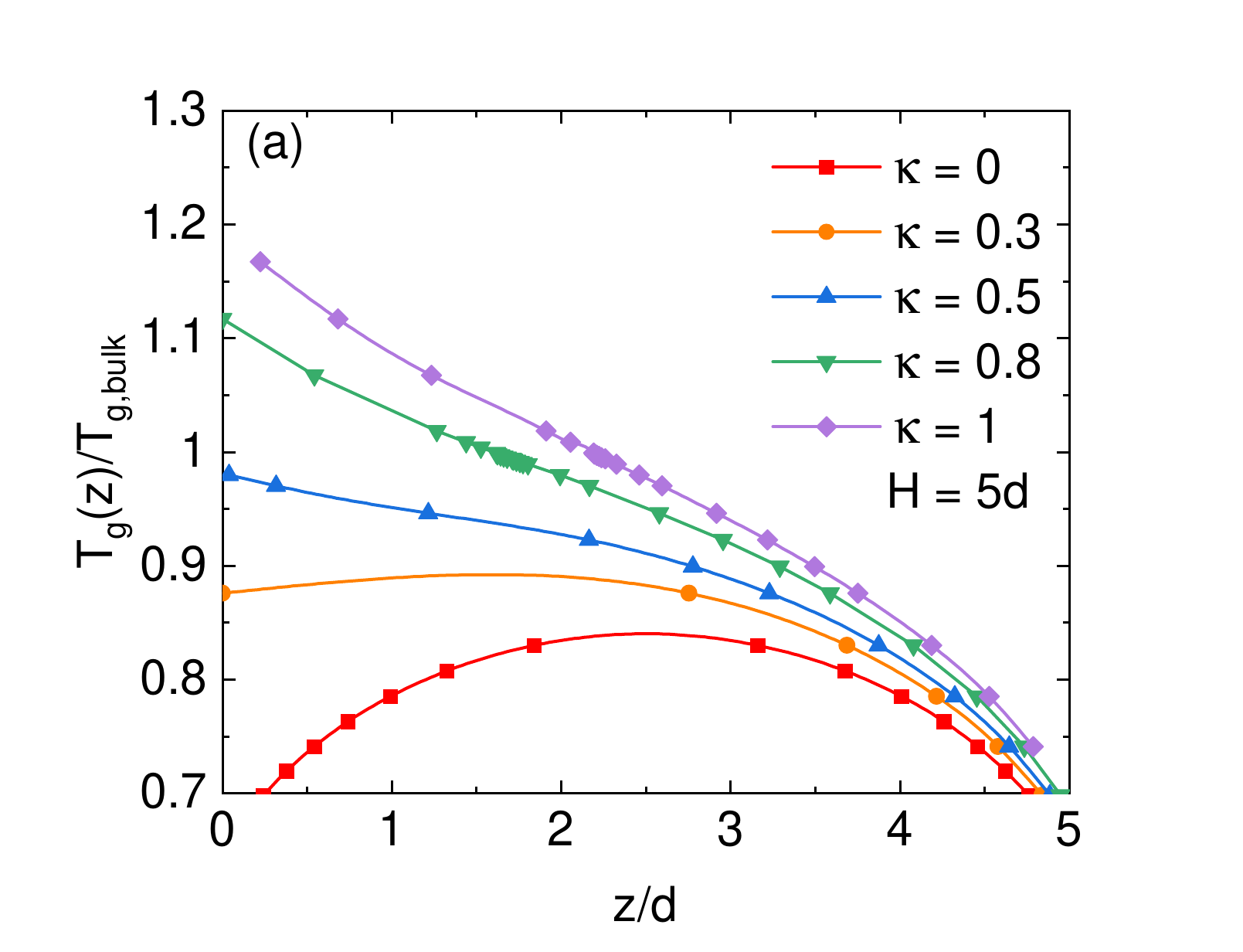}\\
\includegraphics[width=8.5cm]{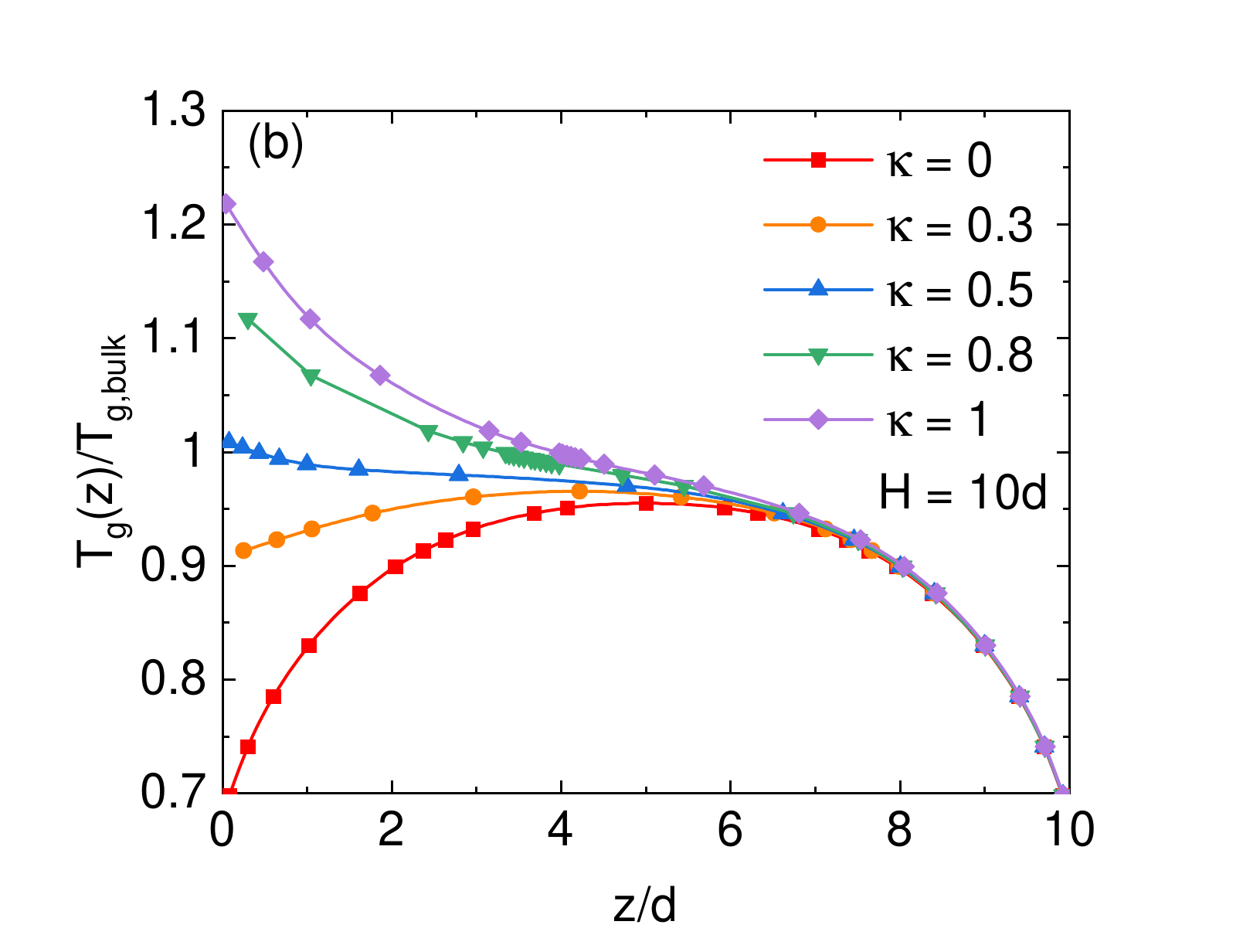}\\
\includegraphics[width=8.5cm]{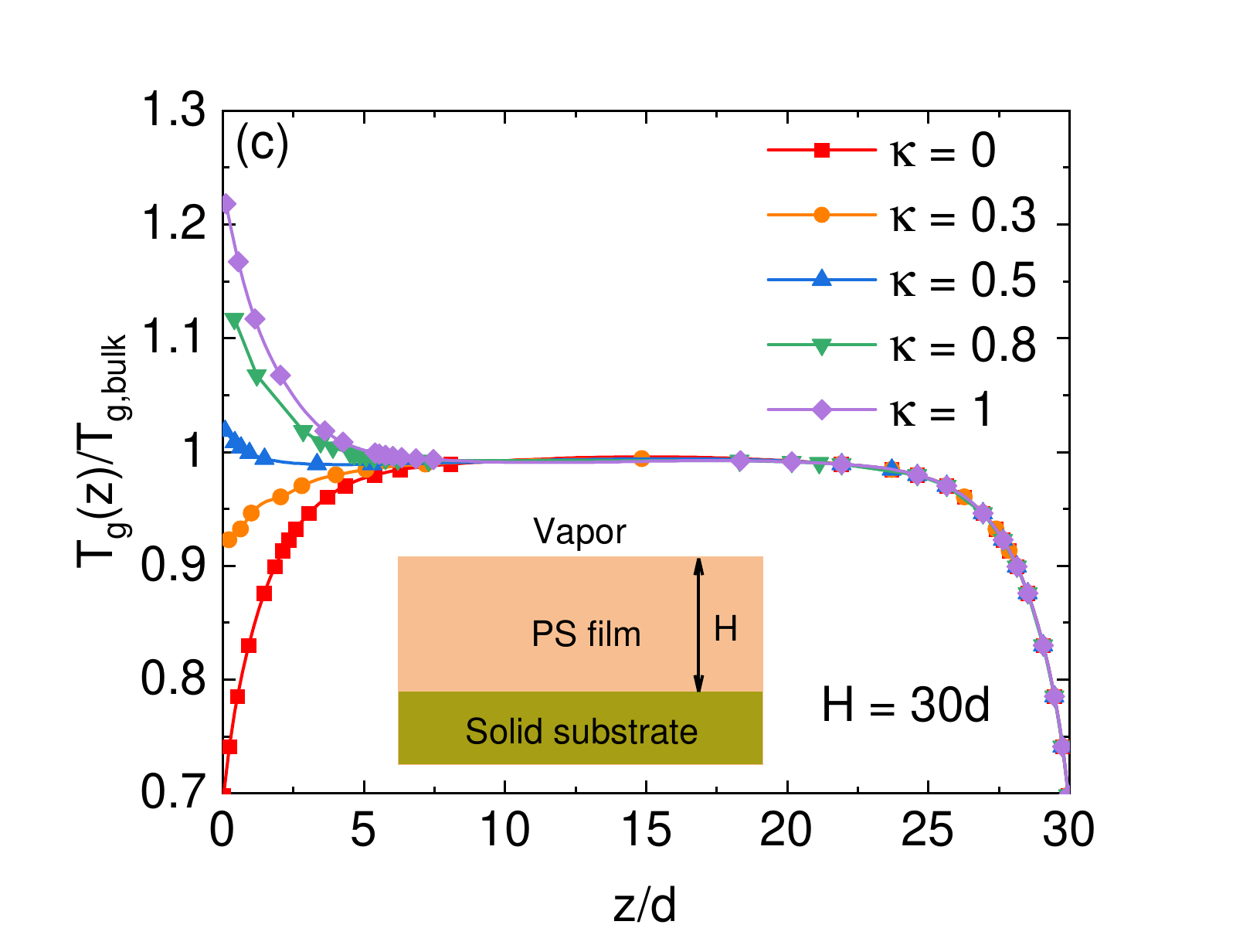}
\caption{\label{fig:7}(Color online) Log-linear plot of the local glass transition temperature normalized by its bulk value as a function of location in a supported PS film for different values of $\kappa$ based on a vitrification criterion of 100 s with film thicknesses of (a) $5d$, (b) $10d$, and (c) $30d$. Here, $T_{g,bulk}$ = 428.3 K corresponds to the typical experimental vitrification time scale of 100 s. For PS, $d \sim 1.1$ nm.} %
\end{figure}

\begin{figure}[htp]
\includegraphics[width=8.5cm]{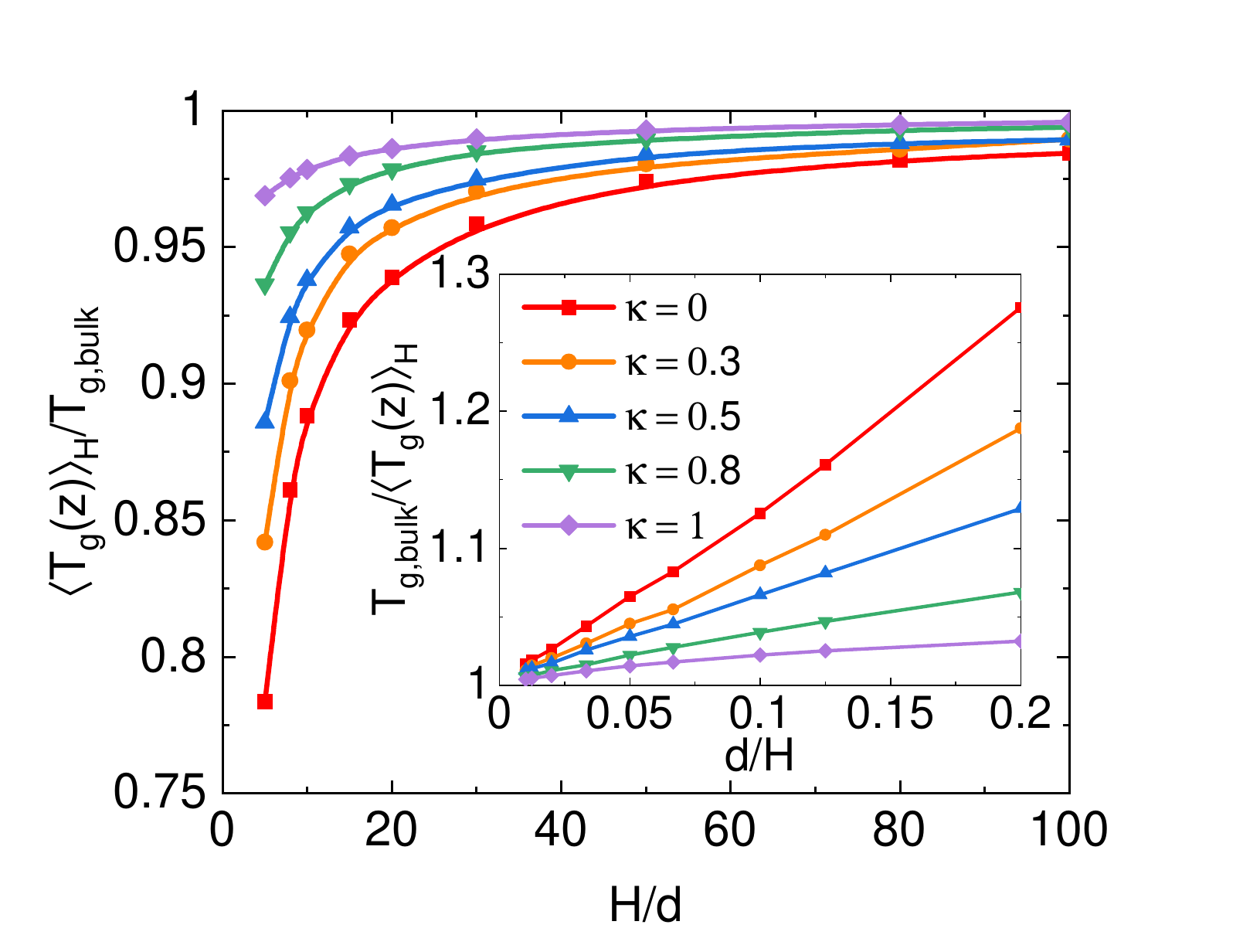}
\caption{\label{fig:8}(Color online) Film-averaged glass transition temperatures (normalized by their bulk value) as a function of film thickness as determined from $\left<T_g(z)\right>_H$ (pseudo-thermodynamic approach) for a vitrification criterion of $\tau_\alpha\left(T_g(z)\right)=100$ s at different values of $\kappa$. Inset: same results as main frame but plotted $T_{g,bulk}/\left<T_g\right>_H$ versus $d/H$.}
\end{figure}

To investigate the dynamic coupling of asymmetric interfacial effects in a thin film, Figure \ref{fig:7} shows the rich variation of the spatial gradient of $T_g(z)/T_{g,bulk}$ for different values of $\kappa$ at various film thicknesses. The presence of the vapor layer ($\kappa=0$) speeds up relaxation, and this acceleration spreads towards the film center. This mechanism accounts for the lower $T_g(z)$ in the film compared to the bulk $T_g$. Increasing $\kappa$ enhances caging at the interface, leading to an elevation of the local glass transition temperature. The maximally rough solid substrate ($\kappa=1$) system exhibits very strongly slowed down dynamics near the interface, and thus $T_g(z) > T_{g,bulk}$. When the film is sufficiently thin ($H\leq 10d$) there is a large separation between the curves in Fig. \ref{fig:8}a which indicates significant interfacial coupling between the two surfaces. Particularly, at the film center ($z=H/2$), one finds $2^{-H/2d}\geq 2^{-5}\approx3.1\%$, and the mathematical forms of Eqs. (\ref{eq:14}) and (\ref{eq:18}) reveal that the effects of both surfaces on the local and collective elastic aspects of the dynamics are important. For $H\geq20d$, the local dynamics recovers its bulk behavior at the film center since  $2^{-H/2d}\leq 2^{-10}\approx0.1\%$ . Thus, we conclude that the behavior of $T_g(z>H/2)/T_{g,bulk}$ is a consequence of only the vapor layer, and the interactions of the two surfaces on the cage scale are dynamically decoupled. Rather, the difference between $T_g(z)$ and $T_{g,bulk}$ arises from the suppression of the elastic barrier associated with the truncation of the displacement field at the interfaces.

As is well known, an effective glass transition temperature of a thin polymer film can be estimated multiple ways \cite{1,6,10,30,31,32,53,54,55}. A thermodynamic-like approach \cite{1,6,30,31,32,54} is based on a "democratic" average over the film, $\left<T_g \right>_H=\frac{1}{H}\int_0^{H}T_g(z)dz$. Such pseudo-thermodynamic calculations can potentially be compared to experimental heat capacity and ellipsometry measurements. Although dynamics near the solid interface is significantly slowed down, particularly if $\kappa=1$, motion is faster near the vapor interface, and can be bulk-like or even faster in the middle of sufficiently thin films. The results in Fig. \ref{fig:8} reveal $\left<T_g \right>_H$ is smaller than its bulk counterpart, and thus in this film-averaged sense the presence of the surfaces enhances the overall molecular mobility. This may seem like a surprising result. But we caution it is not a general result, but rather applies to the present model of a supported film and the parameters that mimic PS. However, it does appear to be qualitatively consistent with experimental findings in Ref. \cite{56} which found the film-averaged glass transition temperature of the supported film is suppressed compared to its bulk value, despite the significant slowing down of dynamics close to a solid substrate. The authors of Ref. \cite{56} speculated this is due to the emergence of a so-called 'dead layer" near the solid surface. In our theory, the detailed mechanism for the average $T_g$ reduction involves the cutoff of collective elasticity at the interface. The specific nature of the substrate-liquid coupling does also play an important quantitative role in the shift of $T_g$, including “dead layers” under some conditions (see Section IV-C), but we find that it is not the primary reason for lowering $T_g$ compared to its bulk value, at least for the model and parameters studied here. For example, for $H=21.5d$ (roughly $H = 25$ nm), we predict $\left<T_g \right>_H/T_{g,bulk}\approx0.966$ for $\kappa=0.5$ and $T_{g,bulk}\sim\left<T_g \right>_H\approx15$ $K$, which, surprisingly and perhaps accidentally, seems to agree quantitatively with experiments in Ref. \cite{56}. Possible caveats to the generality of this conclusion include the model assumption that surfaces do not perturb the structure or density in the film, and the direct effect of strong surface-fluid attractions (adsorbed layers) that are not considered. When $H\rightarrow\infty$, $\left<T_g \right>_H$ recovers the bulk $T_g$, as it must.

\begin{figure}[htp]
\includegraphics[width=8.5cm]{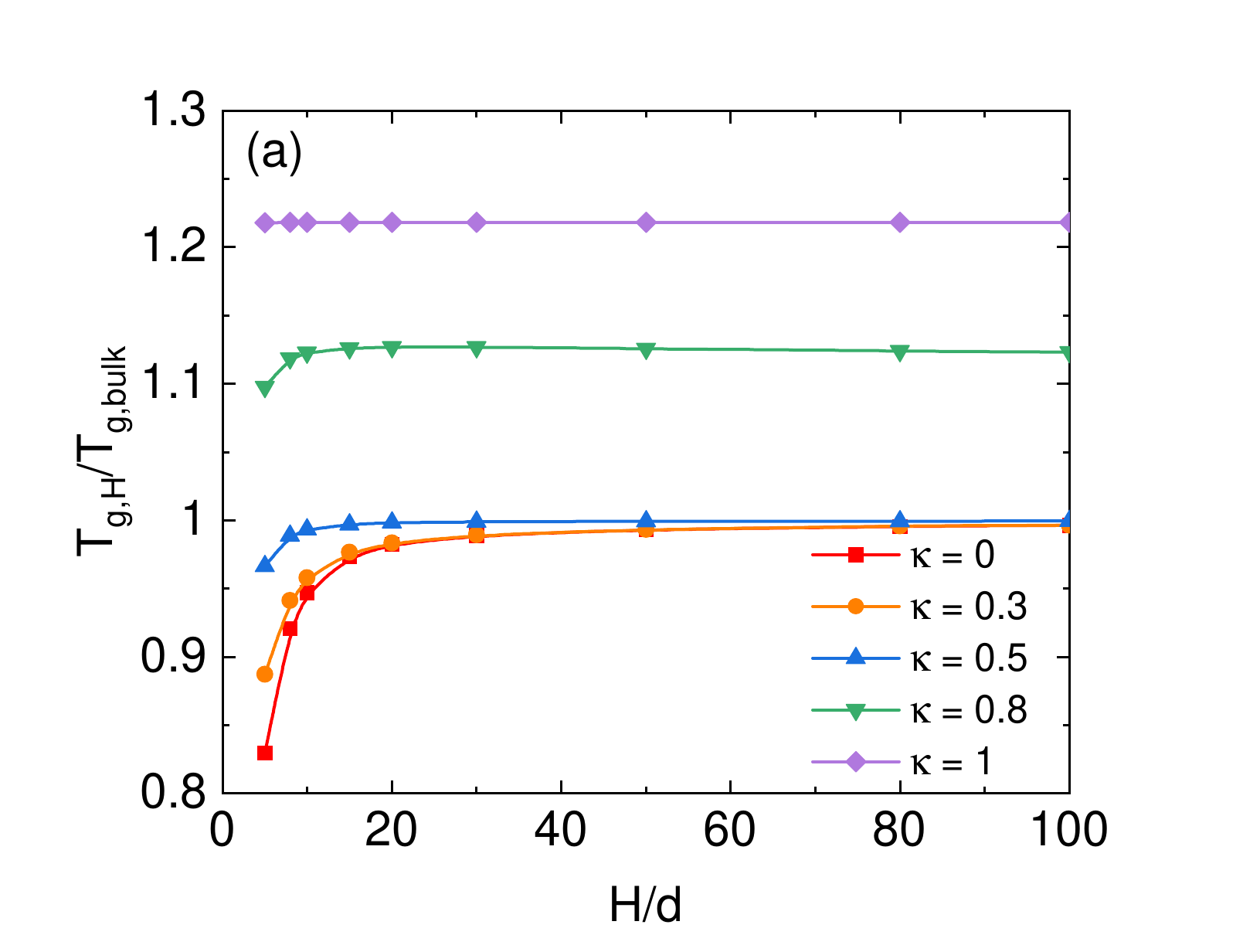}\\
\includegraphics[width=8.5cm]{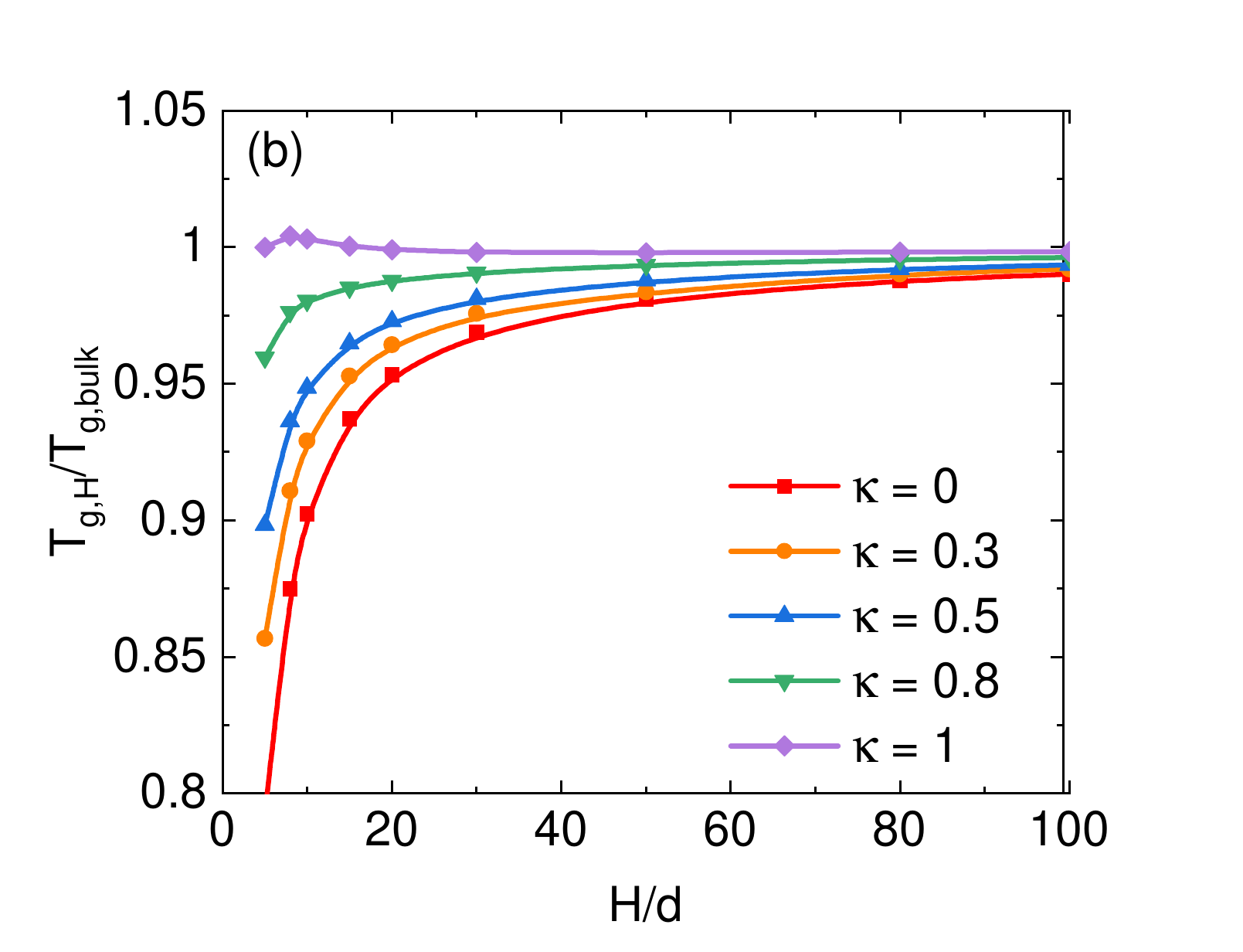}
\caption{\label{fig:9}(Color online) (a) Film-averaged glass transition temperatures (normalized by the bulk analog) as a function of thickness determined using the dynamic criterion $\left<\tau_\alpha\left(T_g,z\right) \right>_H = 100$ s (dynamic approach for the experimental vitrification timescale criterion corresponding to $T_{g,bulk}=428.3$ $K$) at different values of $\kappa$. (b) Analogous results but based on the vitrification criterion $\left<\log_{10}\left[\tau_\alpha\left(T_g,z\right)\right]\right>\equiv2$.}
\end{figure}

As a side comment, one might think that the results in Fig. \ref{fig:8} that suggest the mean film $T_g$ is always suppressed relative to bulk is not consistent with some experiments which measure enhancements in thin films. But the latter are typically materials with significant polymer-surface attractions, which are not in our model. Nor do we allow for attractive surface induced densification near the solid interface. Both these effects would lead to a local elevation of $T_g$, and potentially an increase of the film-averaged vitrification temperature.  Moreover, the results in Fig. \ref{fig:8} do seem natural for our model given the form of the predicted $T_g$ gradients in Fig. \ref{fig:7}c. Of course, if we allowed $\kappa$ to be sufficiently larger than unity, no doubt the film-averaged $T_g$ would be predicted to be larger than its bulk value. We do not consider this parameter regime since, as discussed in Refs. \cite{3,4}, we view $\kappa=1$ as a natural upper bound of the model sketched in Fig. \ref{fig:1}. Finally, we note that in Fig. \ref{fig:9} discussed below, the predicted film-averaged $T_g$ can be larger than the bulk value for high $\kappa$ systems based on a specific gradient averaging criterion, even within our film model.  

An oft-employed empirical expression used to fit experimental and simulation data for how $T_g$ shifts depend on film thickness is based on a naive "two-layer" model \cite{1,10,30,40,41}:
\begin{eqnarray}
\left<T_g(z)\right>_H = \frac{T_{g,bulk}}{1+\cfrac{\delta}{H}},
\label{eq:27}
\end{eqnarray}
where $\delta$ is an adjustable parameter characterizing the interfacial regime. Of course, our theory does not predict the alpha time or $T_g$ gradient is of a step function form corresponding to constant values in two layers, but rather predicts continuous gradients. Moreover, it has been previously shown that the typically valid Taylor series expansion of Eq. (\ref{eq:27}), $\left<T_g(z)\right>_H/T_{g,bulk}=1-\delta/H$, must hold as a mathematical fact for any film for which the dynamic gradient is of finite range \cite{1}. This latter equation provides a good description for the thickness variation of the film-averaged glass transition temperatures irrespective of the specific details of the gradient \cite{1}.

Motivated by Eq. (\ref{eq:27}), we plot our numerical results for $T_{g,bulk}/\left<T_g(z)\right>_H$ as a function of $d/H$ in the inset of Fig. \ref{fig:8}. We find the inverse film-averaged glass transition temperature for diverse solid surfaces is roughly proportional to $1/H$. This linearity validates the empirical usefulness of Eq. (\ref{eq:27}). Note that the slope of the curves increases with decreasing $\kappa$, implying that as the caging constraint at the interface is softened, the apparent interfacial layer length scale ($\delta$) is shortened. 

An alternative approach to determining the effective glass transition temperature of a finite-size film is explicitly based on dynamics \cite{1,10,30,31,32,53,54,55}. One can define a dynamic glass transition temperature as the temperature at which the film-averaged alpha relaxation time $\left(\left<\tau_\alpha(T,z)\right>_H=\frac{1}{H}\int_0^H\tau_\alpha(T,z)dz \right)$ reaches a selected vitrification criterion relevant to dielectric spectroscopy or other dynamic measurements. Figure \ref{fig:9}a shows representative results. We note that the corresponding $\left<T_g(z)\right>_H$ based results in Fig. \ref{fig:8} exhibit a more substantial decrease compared to their purely dynamic counterparts. When $\kappa \leq0.5$, the predictions based on the pseudo-thermodynamic and dynamic approaches exhibit qualitative, but not quantitative, agreement.  

On the other hand, for the more rigid substrates of $\kappa=0.8$ and $1$ the behavior is very different. Specifically, recall that we showed that $\left<T_g(z)\right>_H <T_{g,bulk}$ in Fig. \ref{fig:8}, but the effective film-averaged mean value of $T_g$  determined by $\left<\tau_alpha(T_g,z)\right>_H=100$ s in Fig. \ref{fig:9}a is greater than the bulk $T_g$ out to a large film thickness of $H = 100d$. Such behavior might seem unphysical. However, the construction of the theory guarantees the bulk alpha time will be achieved sufficiently far from the interface, and hence the film averaged $T_g$ will equal its bulk value at “large enough” film thicknesses. Hence, this different behavior for the very rigid substrate cases is a consequence of two factors: (i) how the mobility gradient is averaged, and (ii) our assumption in the calculations that the film structure (and hence alpha time) is fully equilibrated on all length scales. Consideration (i) relates to the democratic weighting of the predicted surface-induced large slowing down of the alpha time near the interface versus the more modest (in a relative, not absolute, sense) vapor-induced speeding up of mobility. Consideration (ii) is highly relevant for large values of $\kappa$ since the theory predicts huge increases of the barrier and alpha time near a rigid rough substrate \cite{4}. However, the latter may not be observed in typical experimental measurements that do not perform ultra long-time aging in which case the near surface layers can appear as effectively “dead”, and the alpha process close to the substrate may not contribute the measured observable from which $T_g$ is determined. For example, if $\Phi=0.58$, we find the alpha time in the equilibrated first layer for $\kappa=1$ is predicted to be $10^7$ s, and the corresponding average alpha time for a thick film of $H = 100d$ is still very large, of order $\sim 10^4$ s, corresponding to a “glass” based on averaging of the alpha time gradient approach. Thus, we indeed expect that $H\gg 100d$ is required to recover the bulk $T_g$.

Of course, if the practical situation in experiment or simulation is the near rigid substrate film is not equilibrated, then our calculations for the high values of $\kappa$ in Fig. \ref{fig:9}a are not directly relevant. We have chosen to present them since they are of conceptual interest as indicating what would be expected in a fully equilibrated film. Moreover, we again emphasize that in the $H\rightarrow\infty$ limit, the theory recovers a film-averaged alpha time and glass transition temperature equal to their bulk values. Explicit numerical verification based on the dynamic averaging criterion requires numerical calculations out to ultra-large values of H which are beyond our ability to computationally treat.

Figure \ref{fig:9}b shows results analogous to those of Fig. \ref{fig:9}a but for the different vitrification criterion of $\left<\log_{10}\left[\tau_\alpha\left(T_g,z\right)\right]\right>\equiv2$. The motivation for this calculation is that when $\kappa > 0.5$, the results in Fig. \ref{fig:10}a are dominated by the contribution of the very long alpha times near the solid surface. The alternative criterion acknowledges the fact, as seen from Figure \ref{eq:7}c, that the normalized gradient of glass transition temperatures near the solid surface is roughly of equal but opposite form (absolute magnitude of shift from the bulk value and spatial range) as that at the vapor surface. Thus, one expects from the alternative averaging approach above that the normalized to the bulk $T_g$ film-averaged glass transition temperature will more rapidly approach its limiting value of unity at large film thicknesses, and its shift from the bulk will decrease and nearly vanish as $\kappa$ becomes large and approaches unity. These anticipated behaviors are indeed found in the numerical results in Fig. \ref{fig:9}b. Taken together, the results in Fig. 9 show (as is well known \cite{1,54}) that in a supported film with strong competing spatial gradients, prediction of an average $T_g$ is a subtle issue that depends on the criterion for averaging over the spatially heterogeneous dynamics.
\subsection{Alpha relaxation time gradients}
\begin{figure}[htp]
\includegraphics[width=8.5cm]{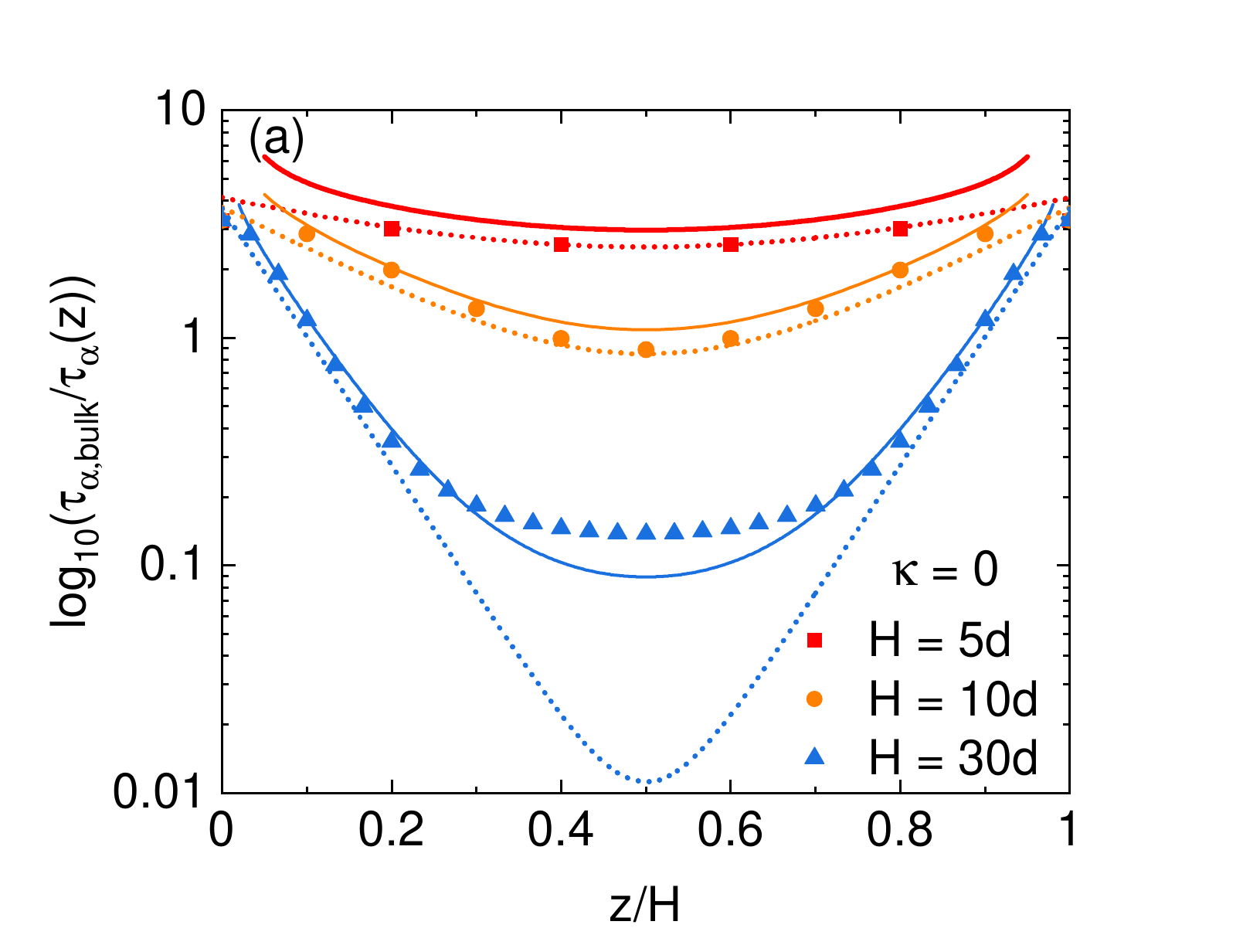}\\
\includegraphics[width=8.5cm]{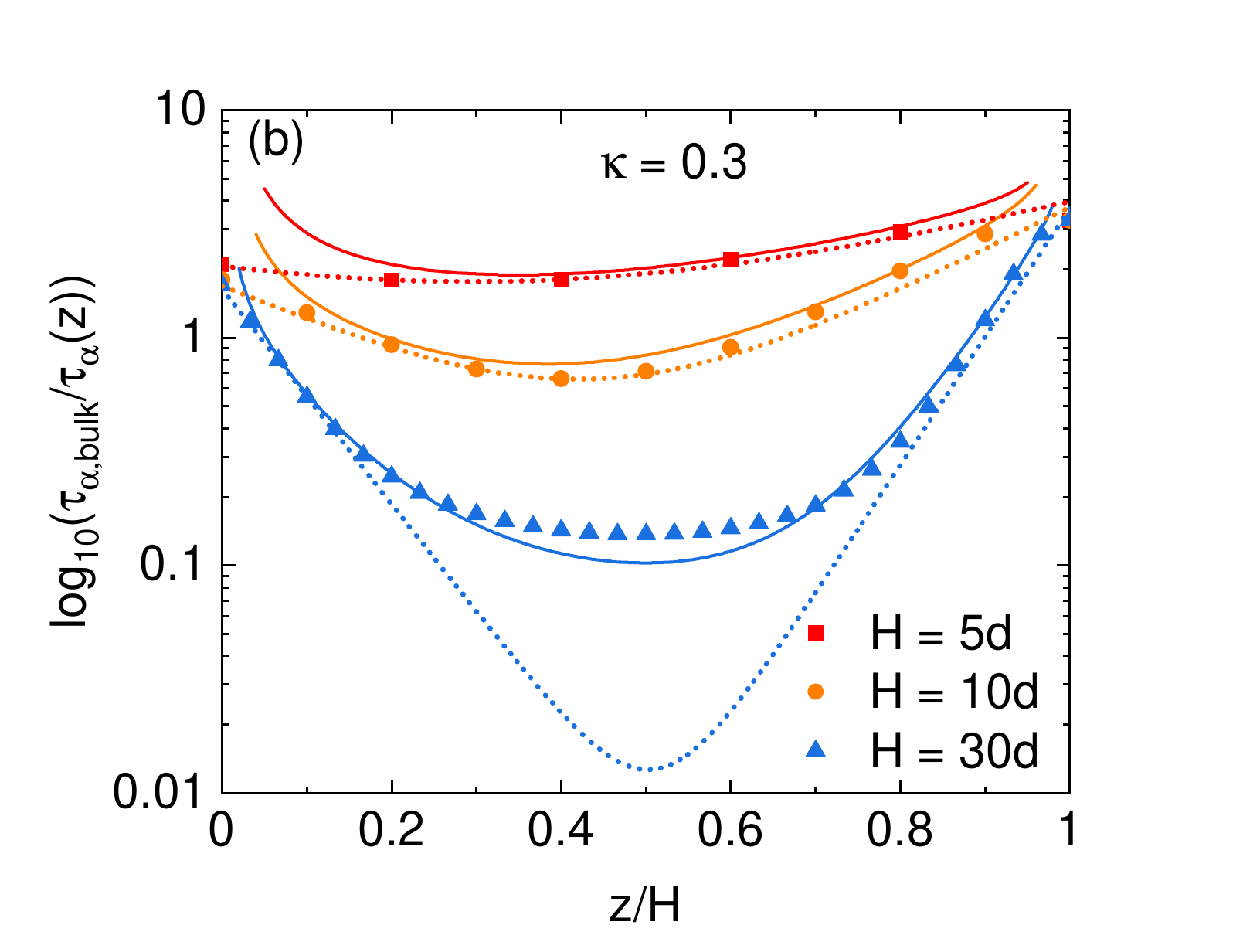}\\
\includegraphics[width=8.5cm]{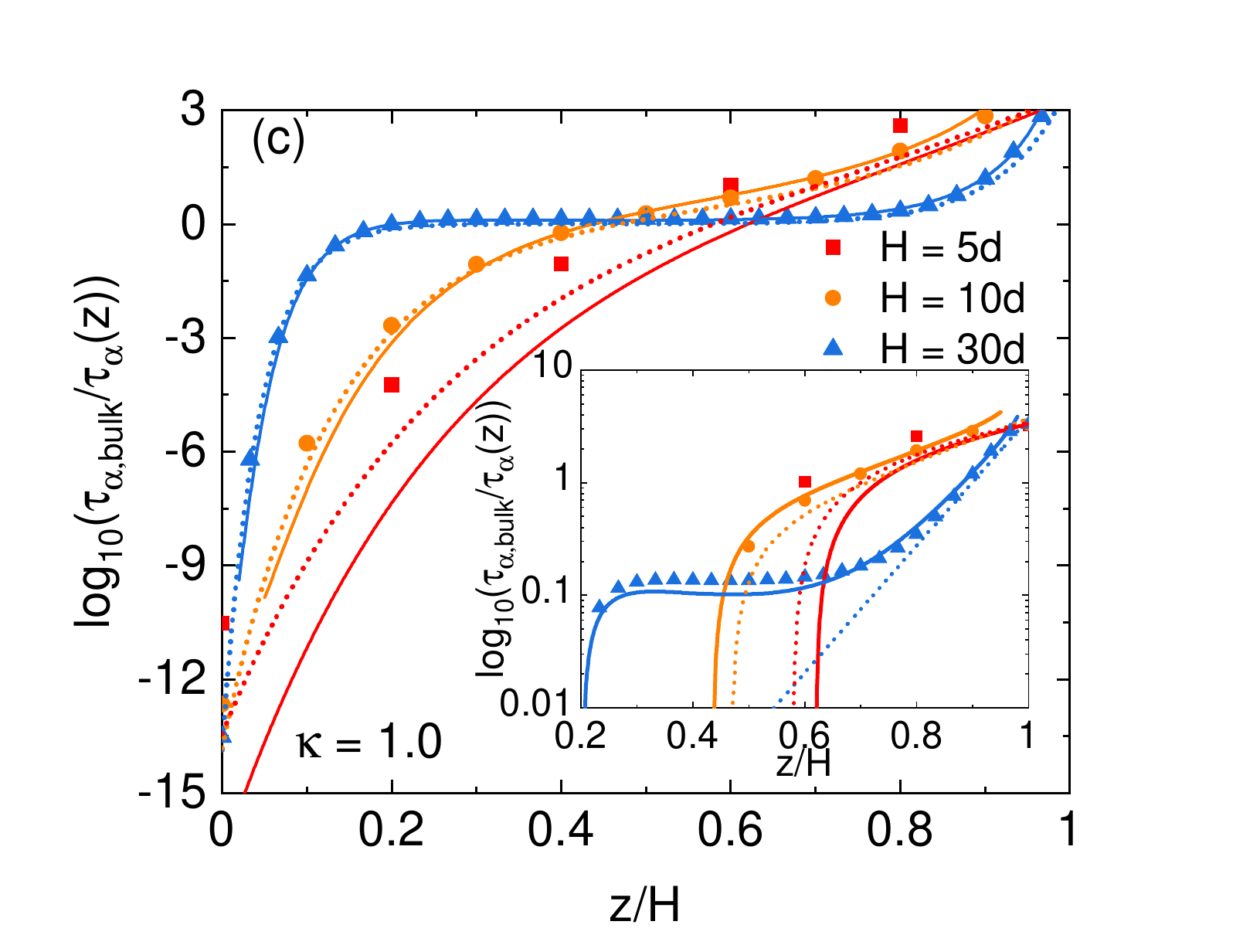}
\caption{\label{fig:10}(Color online) Logarithm of the gradient of the inverse alpha relaxation time normalized by its bulk value in supported films of thicknesses $H= 5d, 10d,$ and $30d$ plotted versus position in the film normalized by the film thickness. Results are shown for (a) $\kappa=0$, (b) $\kappa=0.3$ and (c) $\kappa=1$, at the bulk glass transition of a typical standard MD simulation, which corresponds to a packing fraction of $\Phi = 0.5788$ or a bulk alpha time of $\sim$ 100 ns. Data points and solid curves are the full thin-film numerical calculations and their gradient superposition (Eq. (\ref{eq:28}) analogs, respectively. The dotted curves correspond to the gradient superpositions without the power law decaying term. The inset in Figure \ref{fig:10}(c) plots the same data as in the main frame but in a log-linear format. }
\end{figure}

Numerical ECNLE theory calculations of confinement effects on the alpha time gradient are shown in Fig. \ref{fig:10} and will be discussed in detail below. But we first wish to note that one can attempt to build an analytic description of the gradients as follows. From simulations and our prior analysis \cite{2,7,24}, it is known that the mobility gradient of thin films is sometimes quite well approximated by a linear superposition model corresponding to combining the thick-film gradients of the relaxation time from the solid and vapor interface in a linear manner as \cite{2,7}:
\begin{widetext}
\begin{eqnarray}
\log_{10}\left(\cfrac{\tau_{\alpha,bulk}(T)}{\tau_{\alpha}(T,z,H)}\right) = A_\kappa(T)e^{-z/\xi_\kappa(T)} + \frac{B_\kappa(T)}{z}+A_{\kappa=0}(T)e^{-(H-z)/\xi_{\kappa=0}(T)} + \frac{B_{\kappa=0}(T)}{H-z}
\label{eq:28}
\end{eqnarray}
\end{widetext}
Here, $A_{\kappa=0}(T)$ is the amplitude at the surface, $\xi_{\kappa=0}(T)$ is the decay length (essentially $T$-independent in ECNLE theory [4,5]), and $B_{\kappa=0}(T)$ quantifies the amplitude of the long range collective elastic part of the gradient for $\kappa=0$ which mimics a vapor interface. Adoption of this equation assumes there are no new finite-size effects in the thin-film scenario compared to the thick film. 

Figure \ref{fig:10} compares predictions of the approximate Eq. (\ref{eq:28}) against our full numerical ECNLE theory results. Within the context of the theory, such a comparison provides direct insight concerning the influence of non-additive gradient effects in thin films. This comparison can be done with, or without, the long-range power law contribution to the gradients in order to further probe the importance of the explicitly long-range consequence of collective elasticity on dynamics. 

We now discuss the more detailed trends and features in Fig. 10. Consider first the case of   where our supported film model becomes a symmetric free-standing film as studied in Ref. \cite{2,6}. We find a significant flattening in the mid-film region (Fig. \ref{fig:10}a) for intermediate film thicknesses ($H\approx20-30d$) compared to the gradient superposition approximation without the power law decay term. Considering the power-law tail originates solely from the interfacial truncation of the elastic contribution to the barrier, including it in the analytic superposition expression is expected to improve the accuracy of analytic form.  For a thinner film of $H=10d$, the power-law decay contribution plays a minor role due to the strong cutoff of the elastic field from the two interfaces, and the superposition approximation with and without a $1/z$ power-law tail in $\tau_\alpha(z)$ describe quite well the numerically computed mobility gradients. However, the superposition approximation even with the power law tail gradually becomes less accurate in ultrathin films ($H\sim5d$) due to strong coupling between the different dynamics nucleated at the two interfaces associated with the gradient of cage scale physics. This behavior emerges even though most of the elastic barrier is largely cut-off at the interfaces since the long-range power-law tail disappears in ultra-thin films \cite{2,7}.                           

Figures \ref{fig:10}b and \ref{fig:10}c contrast our full ECNLE theory calculations, analogous superposition approximation results, and calculations based on Eq. (\ref{eq:28}) for the logarithm of the alpha time as a function of $z/H$ for smaller values of $\kappa=0.3$ and $0.5$ and different film thicknesses. The same level of accuracy and limitations of Eq. (\ref{eq:28}) and the superposition approximation found for the $\kappa=0$ free-standing film case are again evident. One sees a good quantitative agreement between numerical ECNLE theory results and the analytic approximation when $H\geq10d$. The inverse power-law decay of the relaxation time gradient is the origin of the gradient flattening in the middle of the film. An important technical limitation of Eq. (\ref{eq:28}) is that it cannot accurately capture glassy dynamics near the interface due to the (artificial) divergence associated with the contributions $1/z$ and $1/(H-z)$. The alpha time gradient only obeys a power law decay form far enough from the surface. In the interfacial region, the analytic formula describing structural relaxation could be much more complicated. 

Given the above analysis, we return to our numerical normalized $T_g$ gradient results discussed above. Motivated by prior work in \cite{7}, we explore whether they can be approximated by the same analytic form as the logarithm of the normalized alpha relaxation time (Eq. (\ref{eq:28})). Thus, we write
\begin{eqnarray}
\frac{T_g(z,H)}{T_{g,bulk}} &=& A_{g,\kappa}e^{-z/\xi_{g,\kappa}} + \frac{B_{g,\kappa}}{z}+A_{g,\kappa=0}e^{-(H-z)/\xi_{g,\kappa=0}} \nonumber\\
&+& \frac{B_{g,\kappa=0}}{H-z},
\label{eq:29}
\end{eqnarray}
where $A_{g,\kappa}, B_{g,\kappa}, \xi_{g,\kappa}, A_{g,\kappa=0}, B_{g,\kappa=0},$ and $\xi_{g,\kappa=0}$ are fit parameters. If we ignore the power decay contributions in Eq. (\ref{eq:29}), the film average $T_g(H)$ becomes
\begin{eqnarray}
\frac{\left<T_g(z,H)\right>_H}{T_{g,bulk}} &=& \frac{A_{g,\kappa}\xi_{g,\kappa}}{H}\left(1-e^{-H/\xi_{g,\kappa}}\right) \nonumber\\ &+& \frac{A_{g,\kappa=0}\xi_{g,\kappa=0}}{H}\left(1-e^{-H/\xi_{g,\kappa=0}}\right).
\label{eq:30}
\end{eqnarray}

\begin{figure}[htp]
\includegraphics[width=8.5cm]{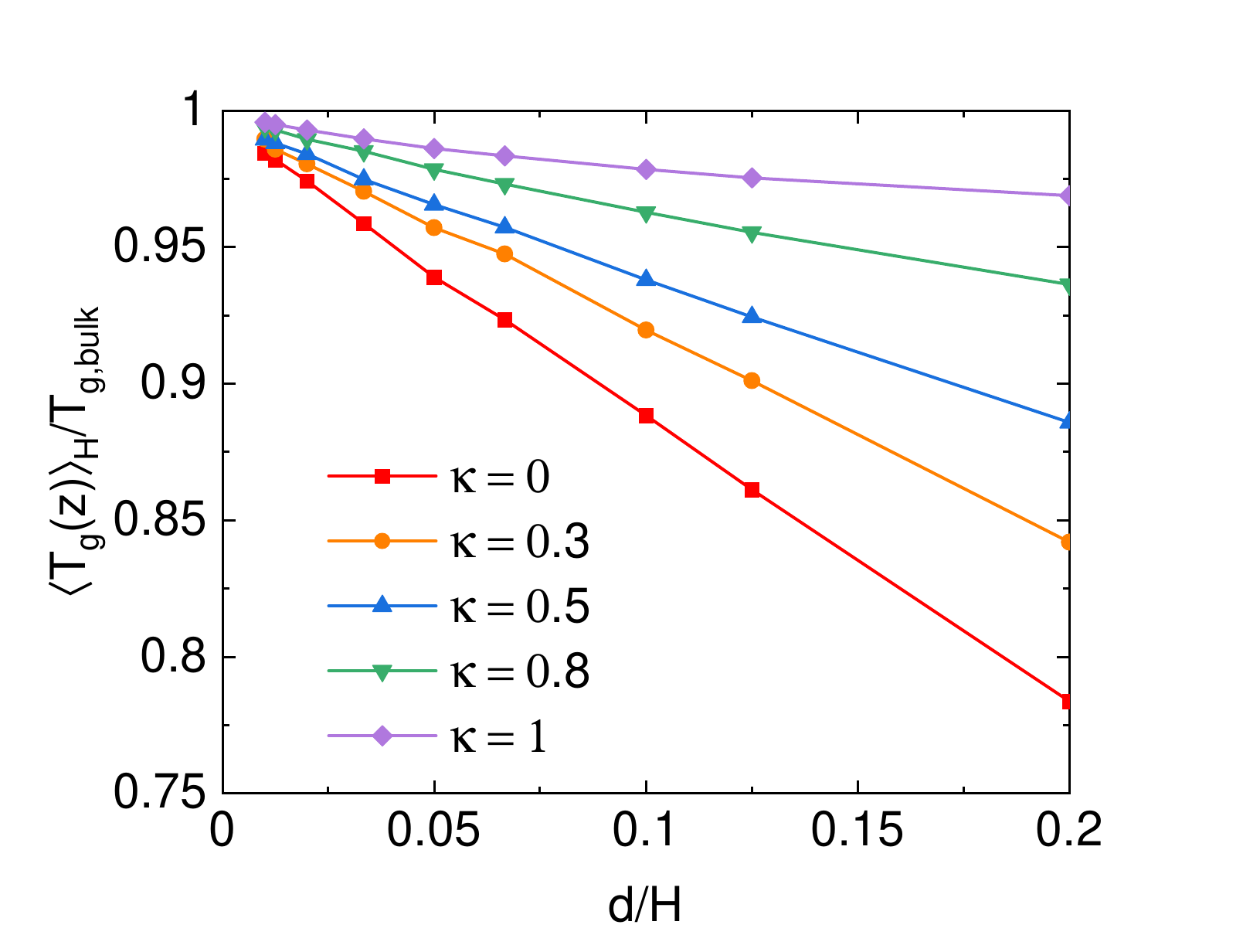}
\caption{\label{fig:11}(Color online) Film-averaged glass transition temperatures normalized by the bulk value of $T_{g,bulk}=428.3$ $K$ (corresponding to the typical experimental vitrification time scale of 100 s) as a function of $d/H$  determined by the pseudo-thermodynamic approach at different values of $\kappa$.}
\end{figure}

Equation (\ref{eq:30}) suggests that $\cfrac{\left<T_g(z,H)\right>_H}{T_{g,bulk}}$ is approximately linearly proportional to $\frac{d}{H}$, particularly when the film is thick. To test this relationship, we replot in Fig. \ref{fig:11} the theoretical data in the main frame of Fig. \ref{fig:8} as $\cfrac{\left<T_g(z,H)\right>_H}{T_{g,bulk}}$ versus $d/H$. We find a good linearity applies for $\kappa= 0, 0.3,$ and 0.5 over a wide range of film thicknesses. However, the film averaged $T_g$ normalized by its bulk value becomes somewhat nonlinear for $\kappa = 0.8$ and 1, particularly when the film is very thin.

\subsection{Glass melting and dead layers in very cold bulk films}
We briefly consider the situation when the bulk liquid is very cold corresponding to being at its experimental bulk glass transition (occurs at $\Phi=0.61$) with an alpha time of 100 s. This situation is particularly germane to the question of the emergence of “dead layers” near a solid interface \cite{10,56,57,58,59,60} (as briefly commented on above), and also to the inverse question of how a vapor interface can lead to melting of a bulk glass which is germane to the formation of so-called ultra-stable glasses. For the former case, the slowing down of relaxation at a solid interface can increase the alpha time beyond what can be measured, and thus in a limited region of space the material is a solid exhibiting a “dead layer”. Of course, in this situation the material in that spatial region is very likely out of equilibrium, so employing equilibrium integral equation theory for the structural input in ECNLE theory calculations is less justified. Our results below should be viewed to within this caveat, but we believe they provide some zeroth order insight to the dead layer question. 

\begin{figure}[htp]
\includegraphics[width=8.5cm]{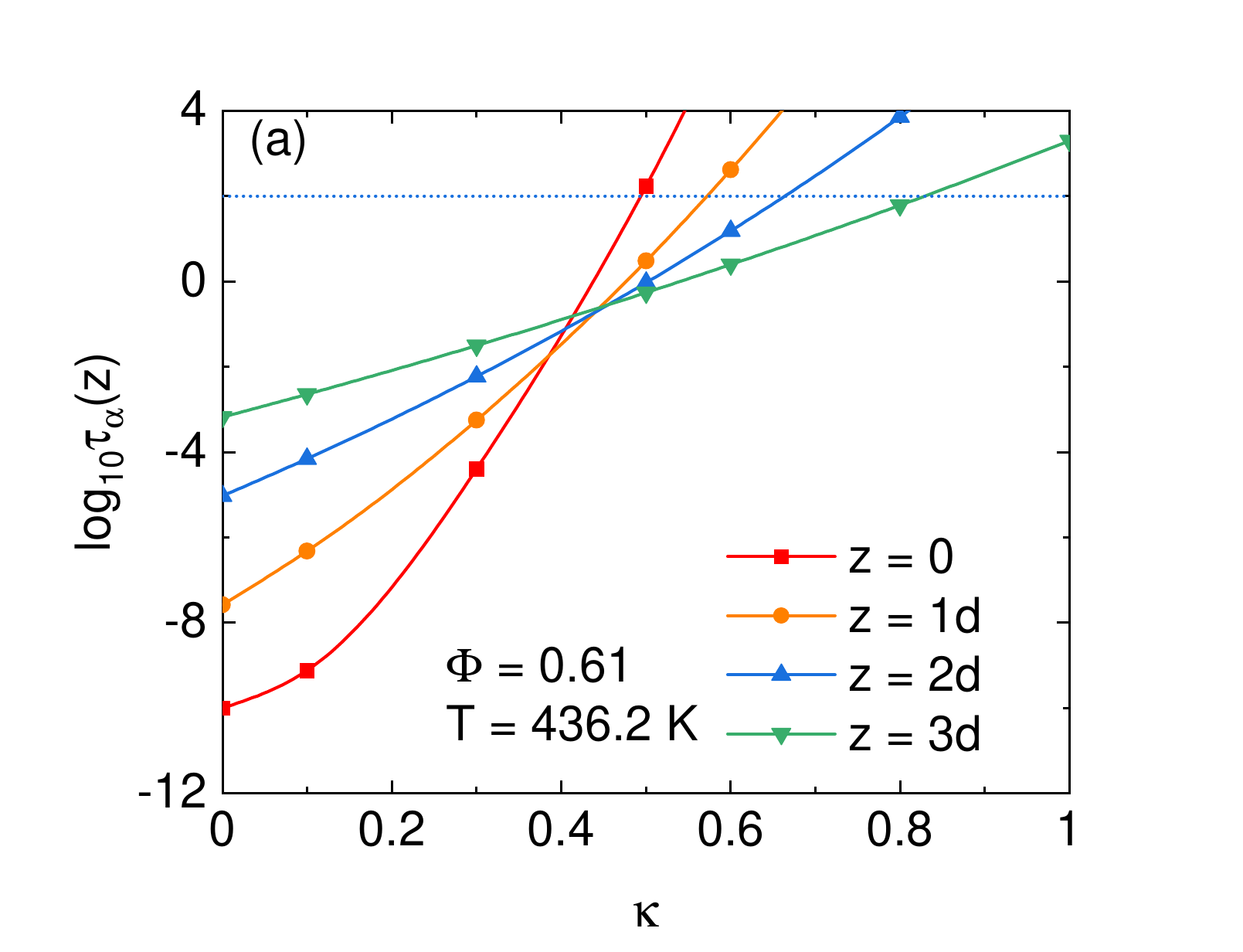}\\
\includegraphics[width=8.5cm]{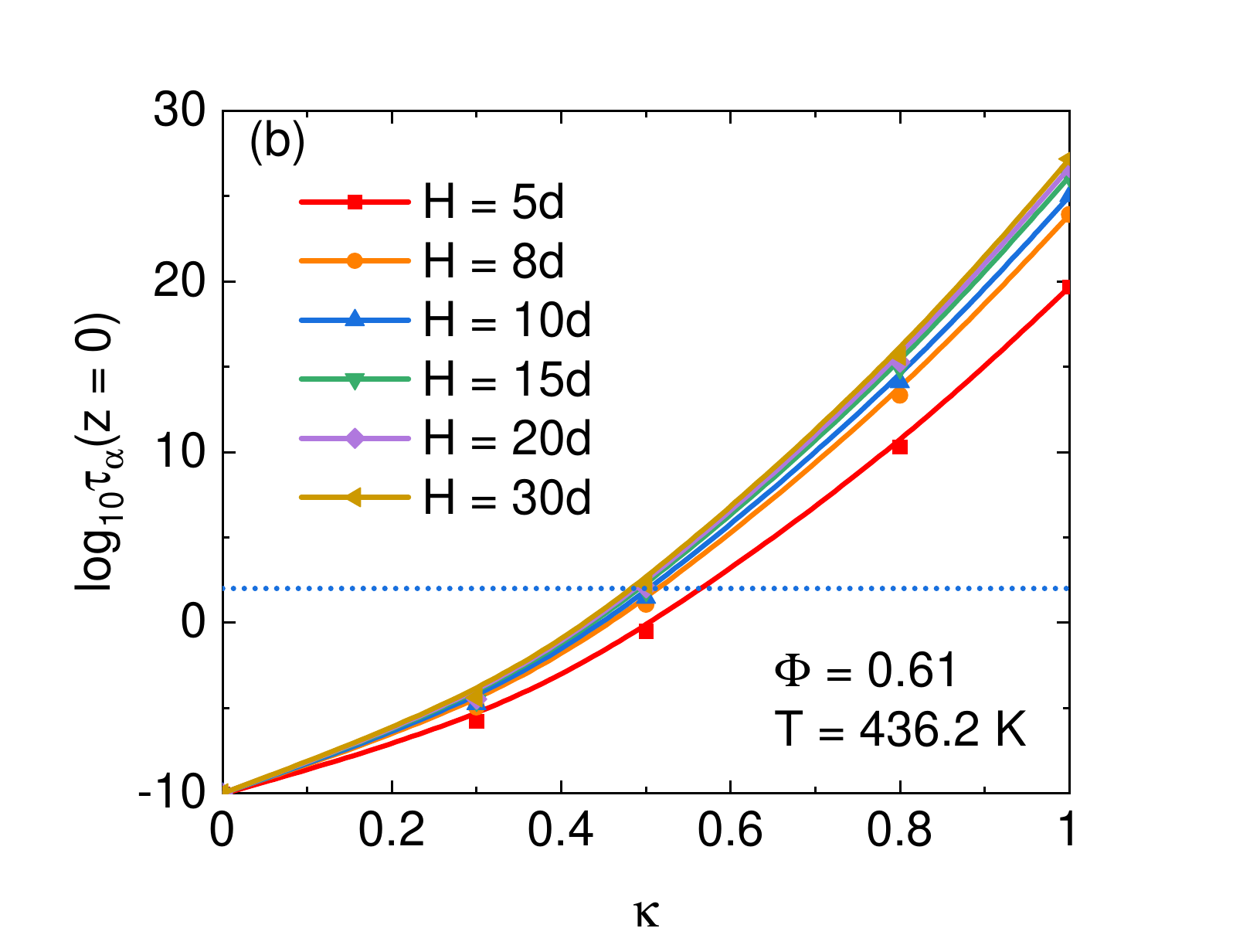}
\caption{\label{fig:12}(Color online) The logarithm of the alpha time (in seconds) calculated at $\Phi=0.61$ or $436.2$ $K$ (bulk laboratory glass transition criterion) as a function of $\kappa$ (a) at positions ($z$) near the solid surface for $H=30d$ and (b) at the solid-liquid interface ($z=0$) for different finite-size films. The results in both panels include the limiting $\kappa=0$ case where the "solid surface" reduces to a vapor interface and hence the system is a free-standing film.}
\end{figure}

Figure \ref{fig:12}a shows the logarithm of the alpha time as a function of $\kappa$ at the PS bulk glass transition temperature at four locations close to the surface in a thick film ($H=30d$). For sufficiently small values of $\kappa$, the bulk glass re-fluidizes, while for larger values of $\kappa$ a dead layer behavior is evident corresponding to an alpha time (far) greater than 100 s. The dead layer is only present for $\kappa > 0.5$, and widens with increasing $\kappa$, extending to beyond 3 particle diameters when $\kappa>0.8$ for the example shown. 

Figure \ref{fig:12}b shows analogous results in the first layer at the solid surface over a wide range of film thicknesses. Now a dead layer emerges not only by increasing $\kappa$ at fixed $H$, but also by increasing film thickness at fixed $\kappa$. This behavior arises from the decoupling of gradients induced by the vapor and solid surface interfaces. Decreasing film thickness facilitates dynamic interference between the gradients emanating from the two interfaces, which reduces the tendency to form a dead layer at the solid surface given the other interface displays faster than bulk dynamics.  A much deeper theoretical analysis of this problem is possible, but is beyond the scope of the present article.

\subsection{Time and frequency domain correlation functions}
The discussion of dead layers in the prior section, and the results in Fig. \ref{fig:9}a that show dramatically enhanced (to high $H$) values of the film-averaged $T_g$ which are likely unobservable due to nonequilibrium effects, are relevant in a fixed temperature sense. However, many experiments near the laboratory bulk glass transition temperature do not directly report a relaxation time, but rather a film-averaged response function which reflects a wide distribution of relaxation times due to strong dynamical gradients. The theoretically predicted massive solid substrate enhancement in relaxation times could lead to a very slow and lower amplitude component of a time domain relaxation function, or a low frequency wing (or perhaps even a distinct peak) in a frequency domain measurement. Such features could be unobservable for practical reasons in typical experiments, and if so, the reported film-averaged mean relaxation time would miss the large substrate induced slowing down effect. This complicated issue would occur even in equilibrium. 

\begin{figure}[htp]
\includegraphics[width=8.5cm]{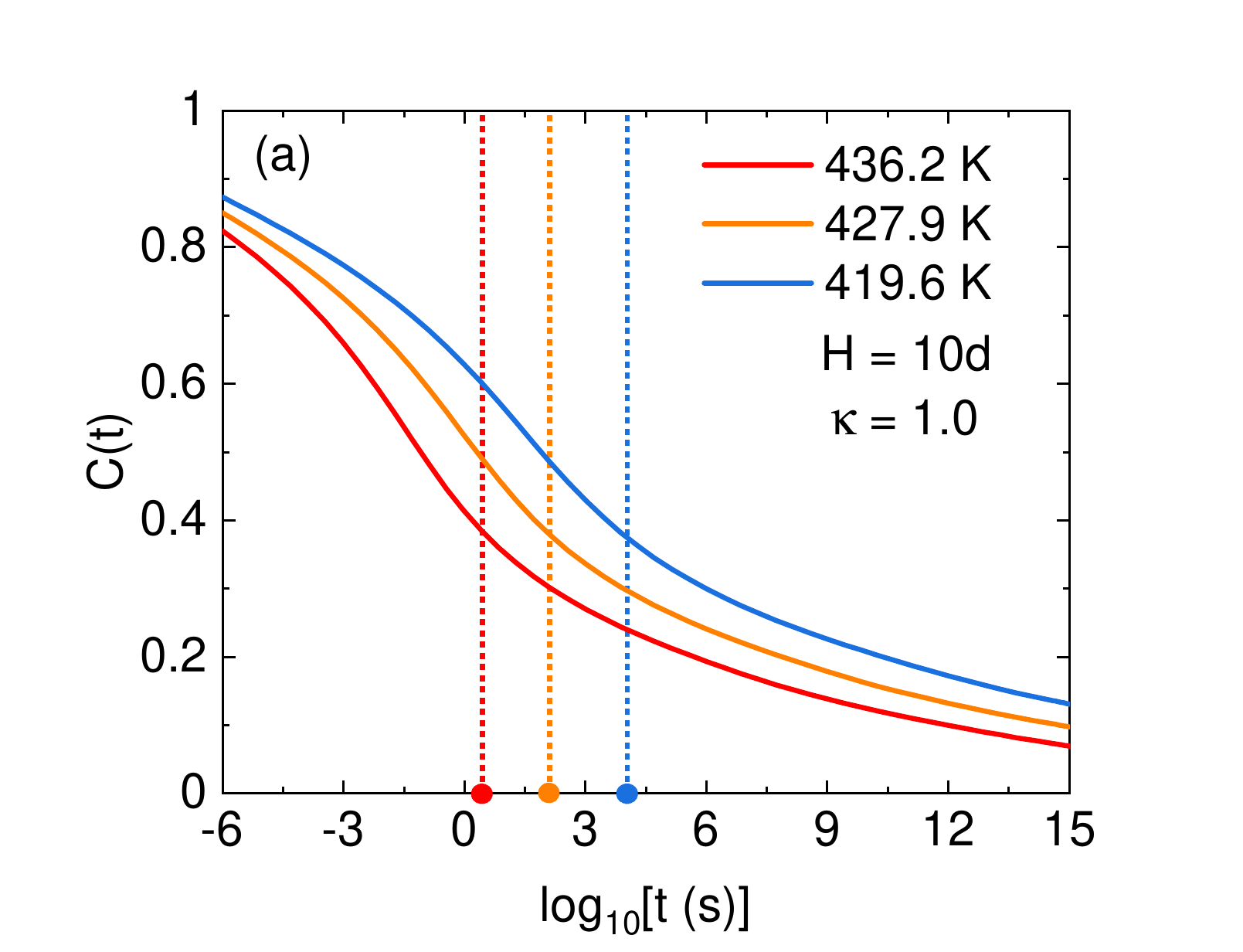}\\
\includegraphics[width=8.5cm]{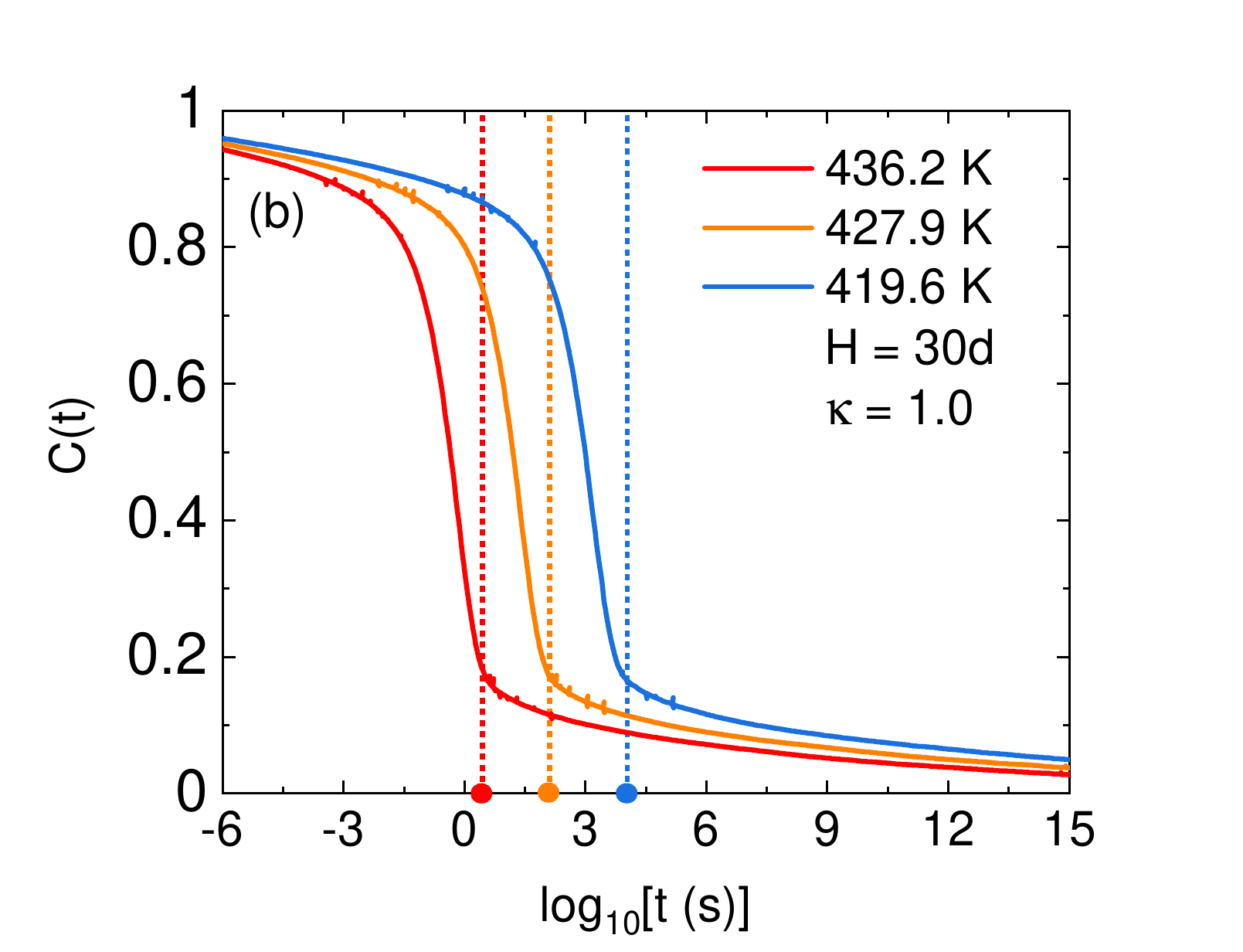}
\caption{\label{fig:13}(Color online) Film-averaged time dependent relaxation function of Eq. (31) for a PS film of thickness (a) $10d$ and (b) $30d$ at temperatures of 436.2 $K$, 427.9 $K$, and 419.6 $K$. The corresponding bulk liquid mean alpha relaxation times expressed as $10^x$ s are $x = 0.441, 2.107$ and $4.03$ with decreasing temperature. The circles on the x-axis indicate the bulk alpha time at the corresponding temperature. The vertical dashed lines are a guide-to-the-eye.}
\end{figure}

\begin{figure}[htp]
\includegraphics[width=8.5cm]{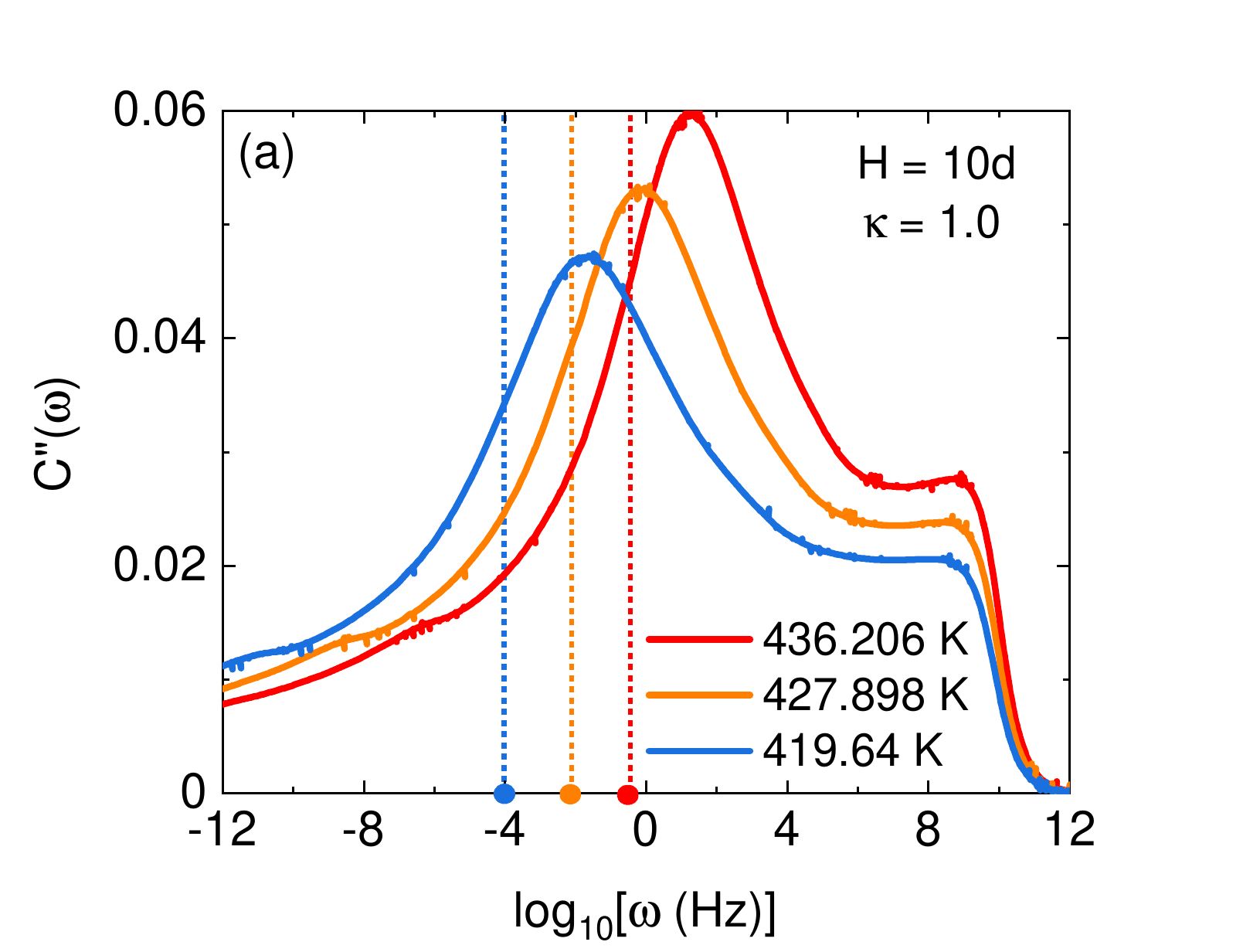}\\
\includegraphics[width=8.5cm]{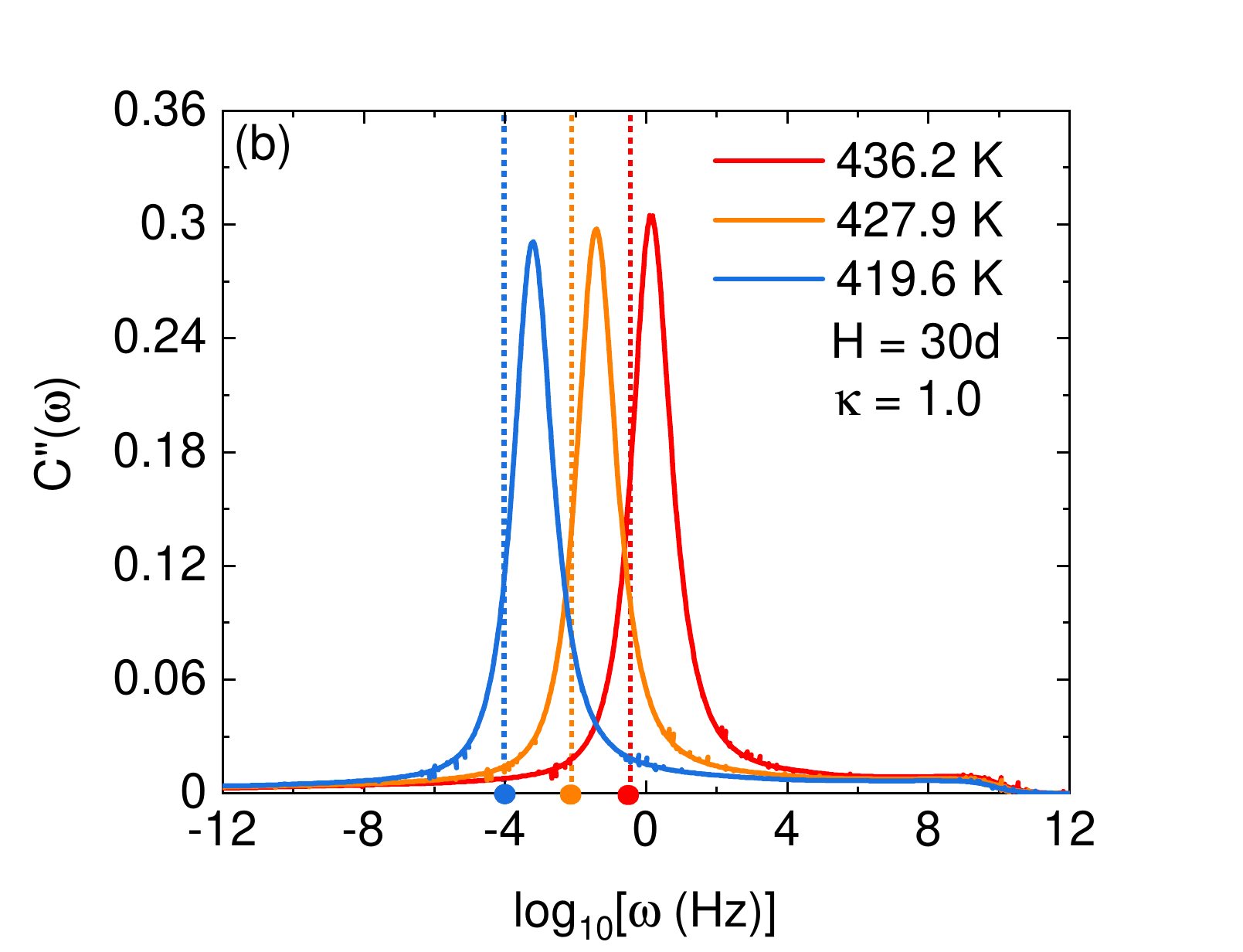}
\caption{\label{fig:14}(Color online) The dielectric loss like function of Eq. (32) at temperatures of 436.2 $K$, 427.9 $K$, and 419.6 $K$ as a function of frequency for PS film thicknesses (a) $10d$ and (b) $30d$. The circles indicate the logarithm of the inverse bulk liquid alpha relaxation time at the corresponding temperatures. The vertical dashed lines are a guide-to-the-eye.}
\end{figure}

The above considerations motivate us to briefly consider what our theory predicts for the film-averaged relaxation function. We ignore any “intrinsic” dynamic heterogeneity that is present even in the bulk, and focus on how interfaces and confinement induce a (spatial) distribution of the alpha time. This corresponds to constructing a film-averaged relaxation function that averages over exponential relaxation processes with time constants that depend on location in the film, i.e., the alpha time gradient considered in this article. Specifically, we define the time and frequency domain film-averaged relaxation functions:

\begin{eqnarray}
C(t) = \frac{1}{H}\int_0^{\infty}dz \exp\left(-\frac{t}{\tau_\alpha(z)} \right),
\label{eq:31}\\
C"(\omega) = \frac{1}{H}\int_0^{\infty}dz\frac{\omega\tau_\alpha(z)}{1+\left[\omega\tau_\alpha(z) \right]^2}.
\label{eq:32}
\end{eqnarray}
Figures \ref{fig:13} and \ref{fig:14} show calculations for the high $\kappa=1$ solid substrate case, two choices of film thickness ($H = 10d$ and $30d$), three temperatures close to, and straddling, the bulk $T_g$ germane to the conditions studied in the previous section and in Fig. 9. The selected temperatures correspond to bulk alpha times of $10^x$ s with $x = 0.441, 2.107$ and $4.03$. There are multiple interesting features.

The time domain results in Fig. \ref{fig:13} display a smeared or continuous two regime form for the thin film ($H = 10d$) where gradients from the vapor and solid surfaces strongly overlap. On the other hand, for the thicker film ($H = 30d$), a well defined, 2-step decay form is predicted. The time constants for a $1/e$ decay of $C(t)$ are close to the bulk alpha time analog for both film thicknesses. For context, we note that the alpha time close to the solid surface at $z = d$ and at the vapor surface are both off scale in Fig. \ref{fig:13}. For the three temperatures shown, the film alpha times expressed as $10^y$s at the solid surface ($z = 1d$) for $H = 10d (30d)$ are given with cooling by $10^y$s with $y$ = 12.2 (13.4), 15 (16.4), and 18.2 (19.8). These timescales are smaller for the thinner film reflecting the significant impact of the vapor interface gradient near the solid surface, and the interface-driven cutoff of the elastic barrier. The corresponding results at the vapor surface are essentially the same for the two film thicknesses, with the alpha times of $10^y$s and exponents of $y$ = 9.98, 9.94 and 9.87 as temperature decreases.

The analogous frequency domain results are shown in Fig. \ref{fig:14}. For each temperature, the primary peak frequency is nearly the same for the $H = 10d$ and $30d$ films, which in turn is nearly identical to that of the corresponding bulk systems. For both film thicknesses, there is a fast (high frequency) component, with a weak peak or shoulder-like feature attributed to the near vapor interface dynamics and truncation of the elastic displacement field.  Its location in frequency space is close to the alpha relaxation rate at the vapor surface. At low frequencies, on the scale shown in Fig. \ref{fig:14}, the thicker film does not exhibit any unusual behavior. In contrast, the thinner film exhibits a broad wing feature extending to ultra-low frequencies that reflects the enormous slowing down of relaxation at the solid surface. Another feature of interest is the peak breadth. For a simple exponential decay process, the response function is of a Lorentzian form with a full width at half maximum (FWHM), $W$, of roughly unity in terms of number of decades. The loss peak of the thick film displays a $W\sim2$ which is essentially invariant to temperature. In contrast, the thinner film peak is enormously broader, with values of $W\sim6-8$ that increase as temperature decreases. 

Overall, these preliminary calculations support the scenario discussed in the first paragraph of this section, and provide a concrete example of the very large effects of dynamical gradients in confined films on experimentally observable film-averaged relaxation functions.

\subsection{Undershoot in the $T_g$ gradient}
As discussed in the two sub-sections above, for $\kappa = 0.8$ and 1.0 the solid interface greatly slows down the interfacial relaxation time compared to its bulk analog, and the alpha time gradient exhibits a steep double exponential decay form near the interface. When the thickness of the supported film is sufficiently large, the impact of the vapor layer on the relaxation of the half of the film near the solid substrate significantly diminishes. The alpha time gradient then follows an inverse power law decay far enough from the interface. The local relaxation time is smaller than its bulk counterpart and progressively increases toward the film center. 

Since the dynamics near the two different interfaces in a supported film are affected in an opposite manner compared to the bulk, and given the long range tail of the gradients associated with collective elastic effects, it is conceivable that an undershoot or overshoot in the normalized mobility gradient in the film might occur under certain conditions. In this section we present calculations that investigate this point. We caution that the very existence of a weak undershoot or overshoot is likely to be a subtle and nonuniversal issue, sensitive to not only the specific model of a supported film and the thermodynamic state studied, but also the technical theoretical approximations adopted and their quantitative consequences for quantitatively predicting the competing gradients at a solid and vapor interface. For example, precisely how the displacement field is constructed (especially the choice of boundary conditions at different interfaces, as discussed in section II), which impacts the quantitative spatial variation of the collective elastic barrier. Such caveats should be kept in mind when considering the significance of the calculations presented below in the context of the validity of ECNLE theory. Indeed, the core ECNLE theory predictions in prior work and the present article are not tied to any possible existence of subtle undershoots or overshoots in mobility gradients.

\begin{figure}[htp]
\includegraphics[width=8.5cm]{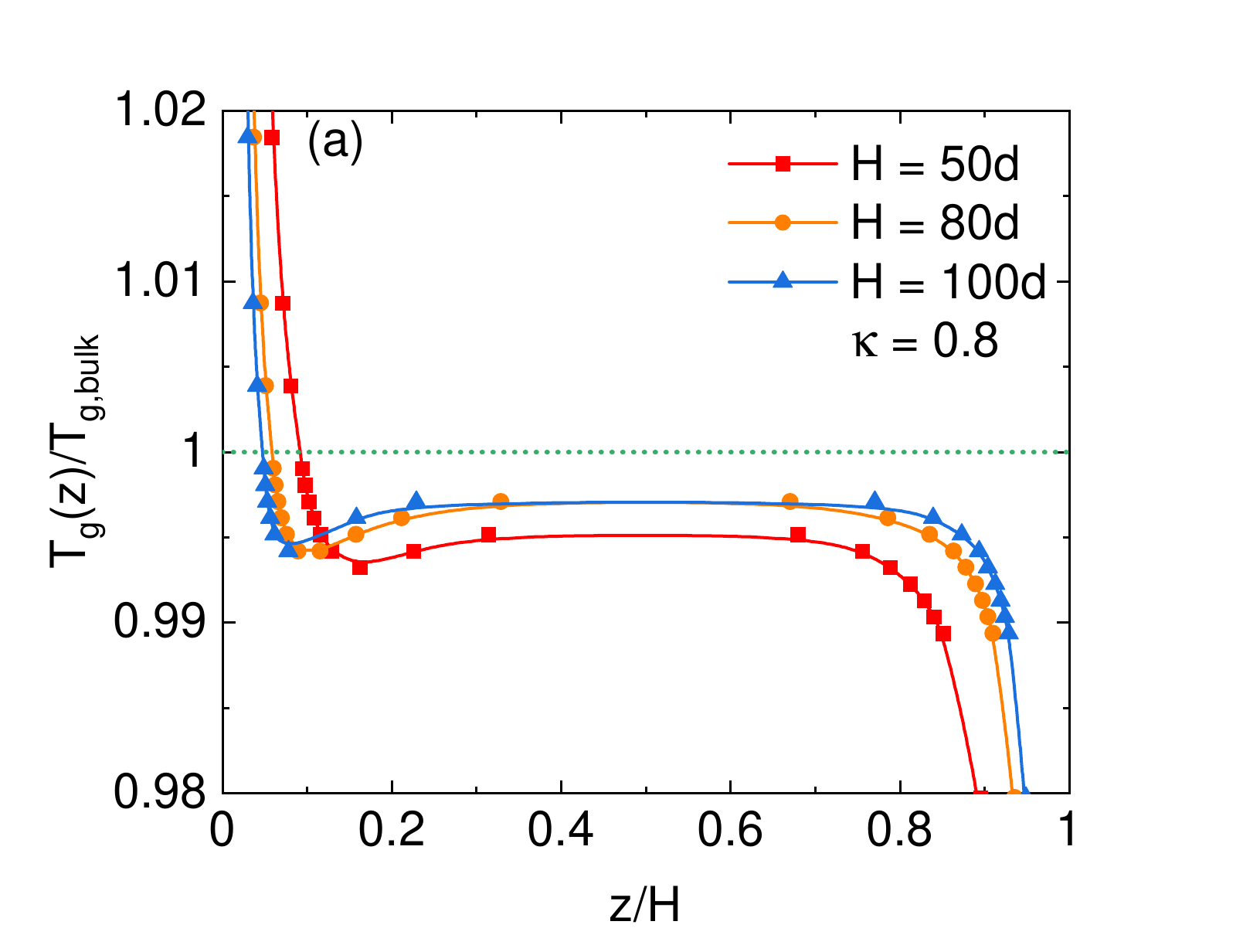}\\
\includegraphics[width=8.5cm]{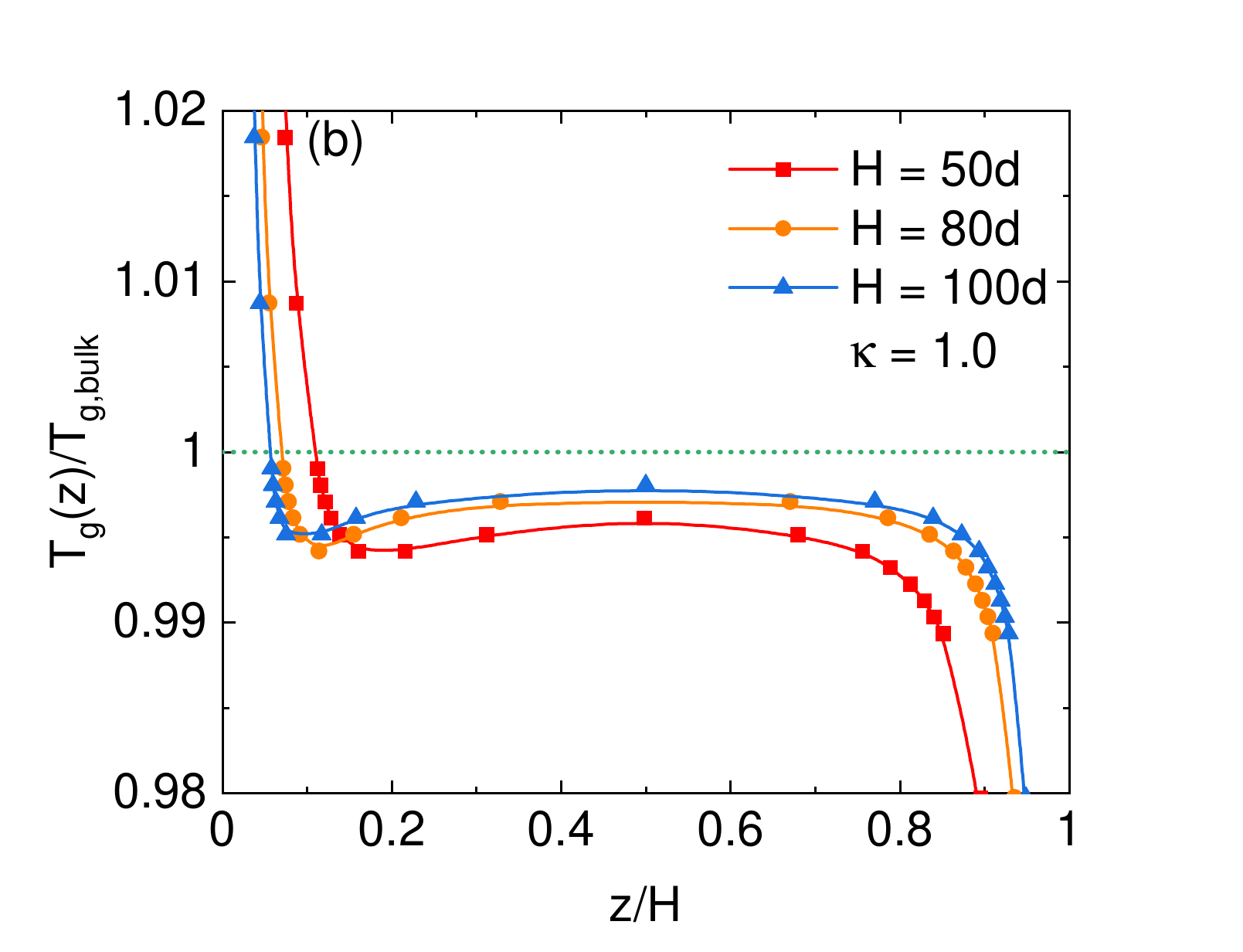}
\caption{\label{fig:15}(Color online) The normalized $T_g$ gradient in supported films of thicknesses $H = 50d, 80d,$ and $100d$ plotted versus position in the film normalized by the film thickness. The results are for (a) $\kappa = 0.8$ and (b) $\kappa = 1$. Here, $T_{g,bulk}=428.3$  $K$ corresponding to the typical experimental vitrification time scale criterion of 100 s.}
\end{figure}

Figure \ref{fig:15} shows calculations of $T_g(z)/T_{g,bul}$ for $\kappa=0.8$ and 1.0 at rather large film thicknesses of $H = 50d$, $80d$, and $100d$. One sees a very low amplitude undershoot (less than a 1 $\%$ effect) at $z/H\sim 0.1$ is predicted. The tiny nature of this effect reinforces the potential concerns expressed above, although the results may also motivate looking for such a feature in simulations. Of course, the latter would be a very challenging task given the small amplitude of this feature. We note that the simulation studies in thick supported films reported in Refs. \cite{15,16} might be suggestive of a weak non-monotonic variation of a mobility gradient, but those results are not definitive for multiple reasons. Another reason for caution concerning the existence of such an undershoot feature is that we expect that when the supported film is "thin enough" the influence of the vapor surface on the dynamics of the entire film becomes sufficiently large that the subtle undershoot feature could disappear. This is indeed what the present theory predicts, as seen from our results for $H=30d$ and thinner supported films in Figs. \ref{fig:7} and \ref{fig:10}, which seems to be consistent with various simulations of polymer films \cite{13,24}. Future theoretical, simulation and experimental work is required to more deeply investigate whether such weak undershoots or overshoots in dynamic gradients truly exist. 

\section{Summary and Discussion}
We have extended ECNLE theory to analyze the gradients of the alpha relaxation time and glass transition temperature, and the corresponding film-averaged quantities, to the geometrically asymmetric case of finite thickness supported films of variable substrate-fluid coupling. The latter involves a complex surface science problem and can be very material specific. In principle, a substrate can either slow down or speed up the dynamics at the interface relative to the bulk behavior due modification of the caging constraints in the dynamic free energy which nucleates a mobility gradient that is spatially transferred into the film. When the film is sufficiently thin, the spatially heterogeneous interfacial relaxation induced by the solid and vapor interfaces are dynamically coupled at both the local cage and longer-range collective elasticity levels. Although increasing the film thickness dramatically reduces the dynamic coupling between the two interfaces with regards to the local dynamics, sharp interfaces of any kind truncate the elastic barrier and lead to generic long-range power law tails in dynamic gradients. Near the interface, the alpha time gradient can be described by a double-exponential form, with an amplitude that depends on density, temperature, and $\kappa$ but with an intrinsic decay length scale that is nearly invariant to thermodynamic state. Overall, the continuous mobility gradient can be qualitatively visualized as composed of three regions: near the solid surface, near the vapor surface, and a “flattened” or “bulk-like” region in the film center. The gradient superposition approximation that includes the power law decay term associated with collective elasticity can provide a rather accurate description of the numerical ECNLE theory predictions for the spatial distribution of relaxation time within the film, particularly in the central region of the film, and for films not too thin. The emergence of near substrate dead layers, strong spatial gradient induced broadening of time and frequency dependent relaxation functions, and the prediction of an unusual but very weak and subtle non-monotonic evolution of dynamic gradient in sufficiently thick and cold films, have also been briefly discussed.

We suggest our new results can be tested using simulations that carefully design in the extent of perturbation of the bulk alpha time at a hard substrate by tuning, for example, its corrugation, mechanical stiffness, or introducing appropriate surface-fluid interactions as recently done by Simmons et al. \cite{7,24,51}. Although direct experimental measurement of such dynamical gradients in supported films is very difficult, our results may provide insight to the highly varied behaviors observed for supported polymer films depending on the nature of the fluid-surface interactions.

Another interesting issue is the question of the level of predictive power of the model and theory. This mainly involves gaining a priori insight to the "proper" value of $\kappa$ for specific surface-fluid systems. This might be achieved based on atomistic computational studies of the very local solid-fluid interface, and/or calibration of the theory against a specific experimental measurement sensitive to the near substrate physics that can allow the deduction of this key parameter, which can then be used to make a priori predictions for other dynamical questions or properties. The most obvious route to achieving the latter is to use fine scale simulation to determine the alpha time of the fluid in the first layer against the substrate and require the theory reproduces it by choice of a single number, $\kappa$. Such an “at the surface” local calibration strategy could then endow predictability of the theory for the rest of the dynamical gradient over a wide range of temperatures and film thicknesses.

Our model and theory can potentially be further developed to study more complex systems and issues such as: (i) dynamical gradients of the polymer alpha relaxation time induced by solid particles or fillers in nanocomposites, (ii) the effects of a densified near surface layer on the film dynamics, and (iii) capped films with two solid interfaces of either identical or different levels of fluid-surface coupling as encoded in the parameter(s) $\kappa$. We emphasize that the basic theoretical ideas for the dynamic free energy gradient have been formulated for any geometry. For example, for a non-flat (curved) solid interface, the pinning fraction of particles at the surface is not 0.5 corresponding to half of the dynamical cage, but rather will depend on the radius of curvature of the interface. 

Concerning capped films with two solid surfaces, we note that the analytic formula of Eq. (\ref{eq:26}) would seem to imply that in a “thick enough” film the tail of the elastic barrier gradient would result in slightly faster dynamics in the film center than in the bulk, despite the huge slowing down in the near surface region. Such a possibility may seem counterintuitive, but much future work is required to investigate this point within ECNLE theory since we have not performed full numerical calculations for such systems, and Eq. (\ref{eq:26}) is based on an approximate analytic analysis that naively superimposes asymptotic behaviors. For any such subtle effect, it is unclear if the latter analytic simplification reflects what ECNLE theory really predicts. Moreover, what is predicted for small effect by the present formulation of ECNLE theory could be sensitive to the details of the approximate construction of the elastic displacement fields and boundary condition choice discussed in section IID.

Finally, if attractive substrate-film interactions are present, this will induce a short-range density gradient orthogonal to the interface. Such a consequence of interfacial interactions was previously treated \cite{3} to model the densified layer(s) and can be extended to the present variable substrate-fluid coupling model. Explicit treatment of in-plane spatial variation of the alpha relaxation time, and the explicit consequences of a substrate-fluid attractive forces, remain open challenges.

\begin{acknowledgments}
We acknowledge many enlightening and informative discussion of the issues in this article with Professor David Simmons.  
\end{acknowledgments}

\section*{Data availability}
The data that support the findings of this study are available from the corresponding author upon reasonable request.

\end{document}